\documentclass[twocolumn,floatfix]{aastex631}

\newcommand{\codesnip}[1]{\texttt{#1}}

\usepackage[]{hyperref}
\hypersetup{
  colorlinks   = true, 
  urlcolor     = blue,
  linkcolor    = blue, 
  citecolor   = blue 
}
\usepackage{amsmath}	
\usepackage{xspace}

\DeclareMathOperator*{\argmin}{argmin}
\DeclareMathOperator*{\argmax}{argmax}

\newcommand{\forbion}[2]{[\ion{#1}{#2}]}
\newcommand{\OIIsel}{\forbion{O}{2}~$\lambda\lambda$3726,3729\xspace}
\newcommand{\Hgamma}{H$\gamma$\xspace}
\newcommand{\Hbeta}{H$\beta$\xspace}
\newcommand{\OIIIsel}{\forbion{O}{3}~$\lambda$5007\xspace}
\newcommand{\Halpha}{H$\alpha$\xspace}
\newcommand{\NIIsel}{\forbion{N}{2}~$\lambda$6583\xspace}
\newcommand{\SIIseli}{\forbion{S}{2}~$\lambda$6716\xspace}
\newcommand{\SIIselii}{\forbion{S}{2}~$\lambda$6731\xspace}
\newcommand{\SIIIsel}{\forbion{S}{3}~$\lambda$9531\xspace}
\newcommand{\OIIIaur}{\forbion{O}{3}~$\lambda$4363\xspace}
\newcommand{\NIIaur}{\forbion{N}{2}~$\lambda$5755\xspace}
\newcommand{\SIIIaur}{\forbion{S}{3}~$\lambda$6312\xspace}
\newcommand{\OIIaur}{\forbion{O}{2}~$\lambda$7319\xspace}
\newcommand{\HeIi}{\ion{He}{1}~$\lambda$3820\xspace}
\newcommand{\HeIii}{\ion{He}{1}~$\lambda$4026\xspace}
\newcommand{\HeIiii}{\ion{He}{1}~$\lambda$4471\xspace}
\newcommand{\HeIiv}{\ion{He}{1}~$\lambda$5876\xspace}
\newcommand{\HeIv}{\ion{He}{1}~$\lambda$6678\xspace}
\newcommand{\HeIvi}{\ion{He}{1}~$\lambda$7065\xspace}
\newcommand{\CIIrli}{\ion{C}{2}~$\lambda$4267\xspace}
\newcommand{\NIIrl}{\ion{N}{2}~$\lambda$5680\xspace}
\newcommand{\CIIrlii}{\ion{C}{2}~$\lambda$6578\xspace}
\newcommand{\CIIrliii}{\ion{C}{2}~$\lambda$7236\xspace}

\newcommand{\HII}{\ion{H}{2}\xspace}

\graphicspath{{./}{figures/}}

\begin{document}
\title[Spectrospatial Models for LVM]{Unified Spectrospatial Forward Models: Spatially Continuous Maps of Weak Emission Lines in the Rosette Nebula with SDSS-V LVM}


\author[0000-0001-7641-5235]{Thomas Hilder}
\affiliation{School of Physics and Astronomy, Monash University VIC 3800, Australia}

\author[0000-0003-0174-0564]{Andrew R.\ Casey}
\affiliation{School of Physics and Astronomy, Monash University VIC 3800, Australia}
\affiliation{Center for Computational Astrophysics, Flatiron Institute, 162 Fifth Avenue, New York, NY 10010, USA}

\author[0000-0002-1264-2006]{Julianne J.\ Dalcanton}
\affiliation{Center for Computational Astrophysics, Flatiron Institute, 162 Fifth Avenue, New York, NY 10010, USA}
\affiliation{Department of Astronomy, University of Washington, Box 351580, Seattle, WA 98195, USA}

\author[0000-0001-6551-3091]{Kathryn Kreckel}
\affiliation{Astronomisches Rechen-Institut, Zentrum f\"{u}r Astronomie der Universit\"{a}t Heidelberg, M\"{o}nchhofstra\ss e 12-14, D-69120 Heidelberg, Germany}

\author[0000-0003-2300-8200]{Amelia M.\ Stutz}
\affiliation{Departamento de Astronom\'{i}a, Universidad de Concepci\'{o}n,Casilla 160-C, Concepci\'{o}n, Chile}

\author[0009-0000-3962-103X]{Amrita Singh} 
\affiliation{Departamento de Astronom\'{i}a, Universidad de Chile, Camino del Observatorio 1515, Las Condes, Santiago, Chile}

\author[0000-0003-4218-3944]{Guillermo A.\ Blanc}
\affiliation{Observatories of the Carnegie Institution for Science, 813 Santa Barbara Street, Pasadena, CA 91101, USA} 
\affiliation{Departamento de Astronom\'{i}a, Universidad de Chile, Camino del Observatorio 1515, Las Condes, Santiago, Chile}

\author[0000-0001-6444-9307]{Sebastián F.\ Sánchez}
\affiliation{Instituto de Astronom\'ia, Universidad Nacional Auton\'oma de M\'exico, A.P. 106, Ensenada 22800, BC, Mexico}
\affiliation{Instituto de Astrof\'\i sica de Canarias, La Laguna, Tenerife, E-38200, Spain}

\author[0000-0002-6972-6411]{J. E. M\'endez-Delgado}
\affiliation{Instituto de Astronom\'ia, Universidad Nacional Auton\'oma de M\'exico, Apartado Postal 70-264, Coyoac\'{a}n, 04510, Mexico City, Mexico.}

\author[0000-0002-6561-9002]{Andrew~K.~Saydjari}
\altaffiliation{Hubble Fellow}
\affiliation{Department of Astrophysical Sciences, Princeton University, Princeton, NJ 08544 USA}

\author[0009-0008-2170-7845]{Luciano Vargas-Herrera}
\affiliation{Departamento de Astronom\'{i}a, Universidad de Concepci\'{o}n,Casilla 160-C, Concepci\'{o}n, Chile}

\author[0000-0002-7339-3170]{Niv Drory}
\affiliation{McDonald Observatory, The University of Texas at Austin, 1 University Station, Austin, TX 78712-0259, USA}

\author[0000-0002-3601-133X]{Dmitry Bizyaev}
\affiliation{Apache Point Observatory and New Mexico State
University, P.O. Box 59, Sunspot, NM, 88349-0059, USA}

\author[0000-0003-3526-5052]{Jos\'e G. Fern\'andez-Trincado}
\affiliation{Universidad Cat\'olica del Norte, N\'ucleo UCN en Arqueolog\'ia Gal\'actica - Inst. de Astronom\'ia, Av. Angamos 0610, Antofagasta, Chile}

\author[0000-0001-8600-4798]{Carlos G. Rom\'{a}n-Z\'{u}\~{n}iga}
\affiliation{Instituto de Astronom\'ia, Universidad Nacional Auton\'oma de M\'exico, A.P. 106, Ensenada 22800, BC, Mexico}

\author[0000-0001-9852-1610]{Juna A. Kollmeier}
\affiliation{Observatories of the Carnegie Institution for Science, 813 Santa Barbara Street, Pasadena, CA 91101, USA}
\affiliation{Canadian Institute for Theoretical Astrophysics, University of Toronto, Toronto, ON M5S-98H, Canada}
\affiliation{Canadian Institute for Advanced Research, 661 University Avenue, Suite 505, Toronto, ON M5G 1M1 Canada}

\author[0000-0002-2368-6469]{Evelyn J.\ Johnston}
\affiliation{Instituto de Estudios Astrof\'{\i}sicos, Facultad de Ingenier\'ia y Ciencias, Universidad Diego Portales, Av. Ej\'ercito 441, Santiago, Chile}

\correspondingauthor{Thomas Hilder}
\email{thomas.hilder@monash.edu}


\begin{abstract}

Analyses of IFU data are typically performed on a per-spaxel basis, with each spectrum modelled independently.
For low signal-to-noise (S/N) features such as weak emission lines, estimating properties is difficult and imprecise.
Arbitrary binning schemes boost S/N at the cost of resolution, and risk introducing biases.
We present a general forward-modelling approach that assumes spectra close on the sky are more similar than distant ones, and so can be modelled jointly.
These “spectrospatial’’ models exploit spatial correlation to provide robust inferences, while simultaneously providing continuous predictions of line properties like strength and kinematics across the sky.
Instrumental and calibration systematics are straightforward to include and infer.
The model provides a natural trade-off between spatial resolution and S/N in a data-driven way.
We apply this to Sloan Digital Sky Survey V (SDSS-V) Local Volume Mapper (LVM) data of the Rosette Nebula, producing continuous maps of fluxes and kinematics for Balmer, nebular, and auroral lines, as well as weak \ion{C}{2} and \ion{N}{2} recombination lines, demonstrating the approach across three orders of magnitude in S/N, including in the very low-S/N regime.
The method recovers identical morphologies across different lines tracing similar ionisation volumes, at varying resolutions set by the S/N.
We additionally provide a general framework for building and fitting such models in \codesnip{JAX}, suitable for many applications.
The implementation is fast and memory efficient, scales to large data volumes as in LVM, and can be deployed on hardware accelerators.

\end{abstract}

\section{Introduction} \label{sec:intro}

Disentangling the physical processes that govern galaxy and stellar evolution requires detailed knowledge of the spatial distribution of stars, gas, dust, and dark matter, as well as the structure and kinematics of the interstellar medium (ISM) \citep{sanchez2020,ma2021,sanchez2021,thaina-batista2025}.
This understanding hinges on reliable, spatially resolved emission line diagnostics of ISM temperature, density, metallicity, ionic abundances, and kinematics \citep{kewley2019}.
Accordingly, mapping emission lines is a central goal of modern integral field unit (IFU) surveys \citep{bacon2001,sanchez2012,bundy2014,green2018,foster2021,emsellem2022,drory2024}.

Faint auroral and metal recombination lines provide the most accurate ISM diagnostics \citep{osterbrock2006,draine2011}, but remain challenging to detect.
Gas-phase metallicities are usually inferred from bright collisionally excited lines (CELs).
Because CEL emissivities depend sensitively on electron temperature $T_e$, investigators either rely on indirect empirical calibrations relating strong line ratios to abundances, or measure $T_e$ directly \citep[e.g.][]{blanc2015,kewley2019}.
Auroral lines are faint CELs from high-excitation states that, when combined with strong lines, provide the $T_e$ measurement needed for direct ionic abundances \citep[e.g.][]{perez-montero2017}.
Metal recombination lines (RLs), yet fainter, have emissivities only weakly dependent on $T_e$ and are often considered the ``gold standard'' for abundance determinations \citep[e.g.][]{peimbert2017,kewley2019}.

A longstanding discrepancy exists between CEL- and RL-based abundances, known as the abundance discrepancy problem \citep[ADP; e.g.][]{wyse1942,peimbert1967,peimbert1969,esteban2004,izotov2006,osterbrock2006,garcia-rojas2007a}.
Proposed resolutions invoke spatial variations in ISM conditions such as temperature and density fluctuations \citep{peimbert1967,mendez-delgado2023,mendez-delgado2023a,mendez-delgado2024} or abundance inhomogeneities \citep{torres-peimbert1990,liu2000}.
Spatially resolved maps of auroral lines, and especially metal RLs, would directly test these scenarios.
To date, such RL maps exist only for bright planetary nebulae \citep[e.g.][]{garnett2001,liu2001,tsamis2008,monteiro2018,garcia-rojas2022,gomez-llanos2024}.

The Sloan Digital Sky Survey V \citep[SDSS-V;][]{kollmeier2017,kollmeier2025} Local Volume Mapper (LVM; \citealp{drory2024}) is an ultra-wide-field IFU survey designed to address these challenges.
Mapping auroral lines is a key LVM science goal, with the prospect of resolving the ADP.
Indeed, LVM is tantalisingly close to the depth required to measure the spatial distribution of metal RLs \citep{singh25}, and so to provide the first gold-standard abundance maps in galactic \HII regions.
Extracting these signals from the data at very low S/N remains challenging.

Addressing this difficulty requires rethinking how IFU data are analysed.
Typically, such analyses are performed on a per-spaxel (a spaxel is a spatial pixel; for a fibre-fed IFU, a spaxel corresponds to a single fibre from a single pointing) basis, where each spectrum is analysed independently of all others.
For very weak spectral features this is often untenable at low S/N.
Binning neighbouring (or even \emph{all}) spaxels boosts S/N, but at the cost of resolution and risks introducing biases.
In this regime, stringent calibration is required: small systematics such as variations in flat-fielding, flux/wavelength calibration, and sky subtraction across pointings, spaxels, and cameras (e.g. for multi-channel and multi-spectrograph instruments like the LVM-I \citep{konidaris2024}) can easily dominate the signal of interest.

We address these issues with a joint modelling approach that exploits the fact that nearby spectra are more likely to be similar than those at larger separations.
Similar previous approaches have either been fully parametric \citep[e.g.][]{bouche2015b}, or have performed spatial reconstruction \emph{after} a typical per-spaxel analysis \citep{gonzalez-gaitan2019}.
Our approach is instead a unified forward model that assumes the observed spectra arise from continuous functions of sky position, subsequently corrupted by systematics and noise.
Spatial dependence is encoded with a fast, memory-efficient Gaussian process prior \citep{greengard2022}, yielding models flexible and tractable for scalable inference on large data volumes.

We apply the method to 19 contiguous LVM pointings covering the Rosette Nebula.
We produce spatially continuous maps of fluxes for strong emission lines (SELs) including Balmer lines, \OIIsel, \OIIIsel, \NIIsel, \SIIseli, \SIIselii and \SIIIsel.
We also map kinematics for these lines, including line-of-sight velocities and intrinsic velocity dispersions (corrected for instrumental broadening).
The SELs selected are identical to those presented in the LVM science overview paper \citep{drory2024} for comparison.
We further map fluxes of auroral lines, \ion{He}{1} recombination lines, and metal recombination lines, demonstrating performance at very low S/N.
Using strong lines, we self-calibrate per-pointing and per-spectrograph systematics under the assumption that the true signal varies continuously across the sky.
Finally, we provide an open-source \citep[\codesnip{JAX};][]{jax2018github} implementation of the spectrospatial framework, which is general, extensible, and can be deployed on hardware accelerators.

\section{Spectrospatial Framework}

\begin{figure*}
    \centering
    \includegraphics[width=0.7\textwidth]{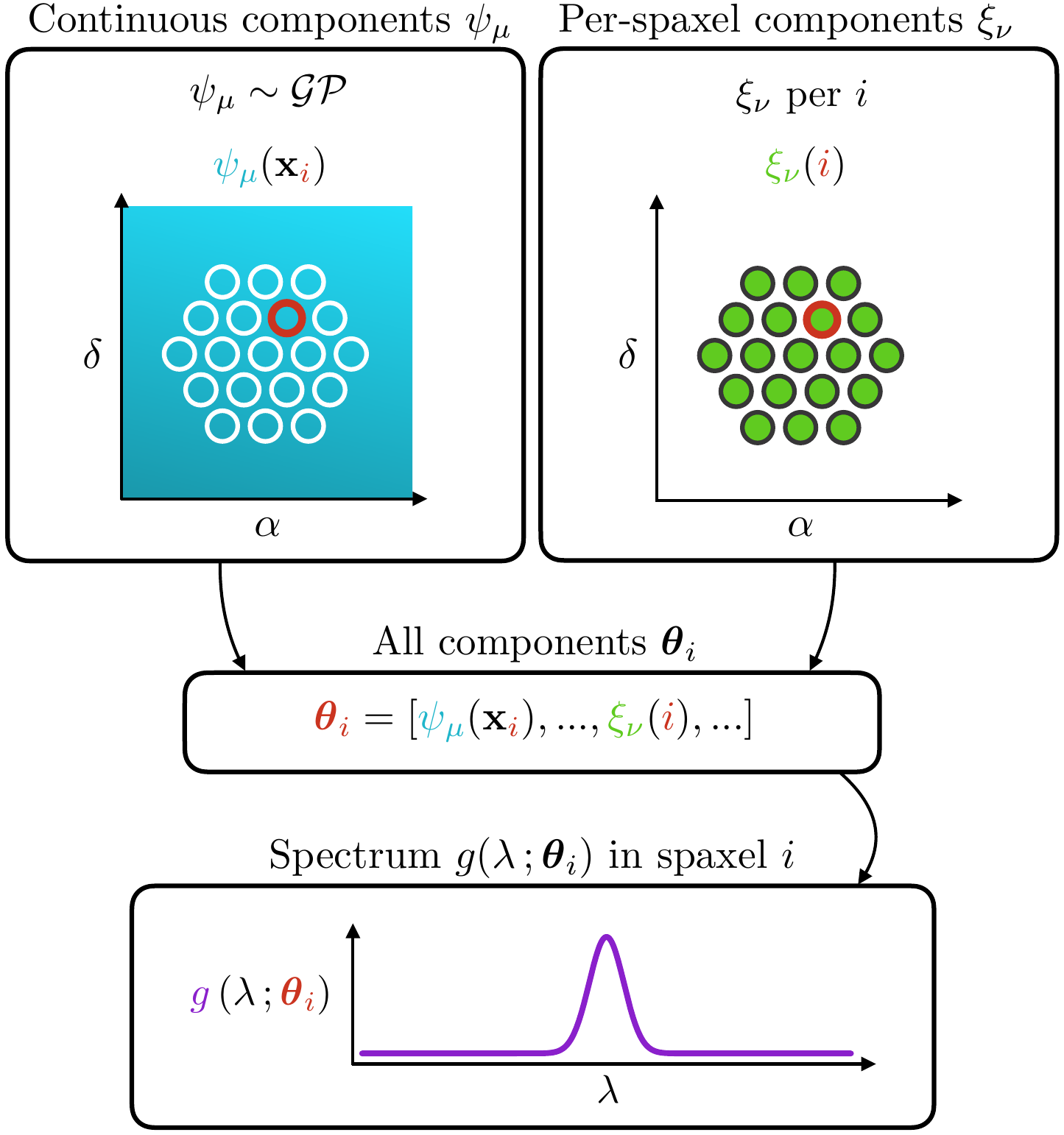}
    \caption{Diagram of the spectrospatial framework. The observed spectrum in spaxel $i$ is generated from a parametric function $g(\lambda; \boldsymbol{\theta}_i)$, where the parameters $\boldsymbol{\theta}_i$ are split into two groups: those that are continuous functions of sky position ${\bf x}_i$ (the $\psi_\mu$), and those that are not (the $\xi_\nu$). The continuous components $\psi_\mu({\bf x})$ are drawn from (transformed) Gaussian processes, while the non-continuous components $\xi_\nu(\cdot)$ are independent per-spaxel $i$, per-pointing $j$, or per-spectrograph $k$. The diagram assumes each $\xi_\nu$ is per-spaxel ($i$) for simplicity. In this paper, we use this framework to model \emph{individual} emission lines, where $g$ is a Gaussian line profile plus a constant continuum component. In general, $g$ could be any parametric function, and could represent multiple lines or even the full spectrum.}
    \label{fig:diagram}
\end{figure*}

Before presenting the implementation and model details for the specific analysis presented in this paper, we first introduce the underlying general framework.
We build up the idea piece-by-piece, culminating in the full spectrospatial forward-modelling framework.
A diagrammatic overview of the framework is shown in Fig.~\ref{fig:diagram}.

\subsection{Motivation and Assumptions} \label{sec:assumptions}

The key motivating idea is that spectra that are close in the sky plane should be more similar than spectra at larger separations.
By this, we specifically mean that the \emph{true underlying spectra} are more similar, not that there is some shared noise that introduces statistical covariance between spectra.
We therefore aim to develop a class of models that is able to leverage the spatially similar information across the spectra.
Ideally, this will allow the model to make more accurate predictions for the properties of the spectra at all points on the sky, as nearby spectra will in some sense ``share'' their information with their neighbours.

We explicitly outline the assumptions adopted in the framework as follows:
\begin{itemize}
    \item \textbf{Parametric spectra:} The spectra in all spaxels are described by a common parametric function $g(\lambda; \boldsymbol{\theta})$, where $\boldsymbol{\theta}$ are the parameters of that function. $g$ does not in general depend linearly on $\boldsymbol{\theta}$.
    \item \textbf{Spatial similarity:} Some parameters of the spectra are not independent per-spaxel, but rather are generated from a shared underlying process that causes spatially nearby spectra to be more similar.
    \item \textbf{Continuity:} The shared spectra properties are continuous functions of sky position.
    \item \textbf{Stationarity:} Our prior belief in the degree of similarity between spectra depends only on the distance between them, not their absolute position on the sky.
    \item \textbf{Noise independence:} The spectra are statistically independent across different spaxels, in that the noise processes affecting each spaxel are uncorrelated.
    \item \textbf{Gaussian noise:} The noise in all pixels is zero-mean, additive and Gaussian, with known variance given by the measurement uncertainty.
    \item \textbf{MAP $\boldsymbol{\approx}$ GP posterior mean:} The maximum a posteriori (MAP) estimate of the Fourier representation of the latent Gaussian processes (GPs) components approximates the posterior mean of the GPs.
\end{itemize}
Each of these assumptions is discussed and built upon in the following sections.

\subsection{Forward Models} \label{sec:forward_models}

The general idea of forward modelling is to construct a model of the underlying data generation process itself, and then to use the data we actually observe to infer the parameters of that model \citep[e.g.][]{gelman2013}.
For the spectrum observed in a single spaxel $i$, we write this as
\begin{align}
    d_i (\lambda_j) \sim \mathcal{N} \left( g (\lambda_j \, ; {\boldsymbol \theta}_{i} ), \sigma_{ij}^2 \right), \label{eq:forward_spaxel}
\end{align}
where $d_i (\lambda_j)$ is the measured spectral flux density at wavelength $\lambda_j$, $g$ is a parametric function that describes the shape of the spectrum, and ${\boldsymbol \theta}_{i} = [\theta_1(i), ..., \theta_\mu(i)]$ is a vector containing the parameters of that function for spaxel $i$.
These parameters might be astrophysical (e.g. the amplitude, velocity shift, and broadening of an emission line) or instrumental (e.g. flux or wavelength calibration systematics).
The $d \sim \mathcal{N} (m, \sigma^2)$ notation means that the observed data $d$ is assumed to be drawn from a Normal (Gaussian) distribution with mean $m$ and variance $\sigma^2$.
So in Eq.~\eqref{eq:forward_spaxel}, we assume the observed spectrum has been generated by random draws from a normal distribution with a mean given by the parametric model $g$ and a variance given by the measurement uncertainty on each data point.

The typical statistical framework for forward models is that of Bayesian inference. 
Bayesian inference provides a mathematical formalism for encoding beliefs about the data generation process, called priors, and a set of instructions for updating those beliefs after observing the data, called Bayes' theorem.
Priors can be written similarly to above, for example
\begin{align}
    {\boldsymbol \theta}_{i} \sim p ({\bf Z}),
\end{align}
where $p$ is some probability distribution that we place over the parameters of the spectrum $\boldsymbol{\theta}_i$, and ${\bf Z}$ are parameters of that distribution.
This idea is critical for our approach, as it will allow us to incorporate our belief that spatially nearby spectra should be more similar.

\subsection{Latent Gaussian Processes}

To extend the per-spaxel forward model we wrote in Eq.~\eqref{eq:forward_spaxel}, to a spectrospatial forward model over \emph{all} spaxels, we first modify our per-spaxel $i$ spectrum parameters $\boldsymbol{\theta}_i$ to instead be continuous functions of sky position ${\bf x}_i = [\alpha_i, \delta_i]$, where $\alpha_i$ and $\delta_i$ are the right ascension and declination of spaxel $i$.
We write 
\begin{align}
    \boldsymbol{\theta}_i = \left[ \theta_1 ({\bf x}_i), ..., \theta_\mu ({\bf x}_i) \right],
\end{align}
where each $\theta_\mu$ is the function for component $\mu$, and these components are still the properties of the spectrum function $g$, for example the height, centre or width of an emission line.
Replacing discrete, per-spaxel values with continuous functions will allow us to model the spatial dependence of the properties of the spectra, and to encode our belief about the similarity between nearby spectra.
We will refer to the functions $\theta_\mu({\bf x}_i)$ that replace the parameters $\theta_\mu(i)$ of the spectrum as ``components'' of the spectrum.

In general we may not actually want \emph{all} components to be continuous on the sky, and more similar in nearby spaxels.
We may want some aspects of the spectra to be different in each spaxel $i$, or each pointing $j$, or each spectrograph $k$, and independent of sky position ${\bf x}_i$, for example to account for unknown random variations in flux calibration across pointings.
To account for this, we split the spectrum components into two groups: those that are continuous on the sky called $\psi_\mu({\bf x}_i)$, and those that are not called $\xi_\nu(\cdot)$, where the argument denotes the relevant discrete index ($i$, $j$, or $k$).
Thus
\begin{align}
    \boldsymbol{\theta}_i &= \left[ \psi_1({\bf x}_i), ..., \psi_\mu({\bf x}_i), \xi_1(\cdot), ..., \xi_\nu(\cdot) \right], \\
    &= [\boldsymbol{\psi} ({\bf x}_i), \, \boldsymbol{\xi} (\cdot)],
\end{align}
where the bolded terms are $\mu$- and $\nu$-dimensional vectors collecting the continuous and non-continuous components of the spectrum, respectively.

To model these continuous functions $\psi_\mu$, we use transformed Gaussian Processes (GPs).
We refer the reader to \citet{rasmussen2006} for a comprehensive introduction to GPs, but we briefly summarise the key ideas here.
GPs allow us to place a prior over a space of functions, based on the properties of the kind of functions we think are appropriate for the spectrum parameters.
We write this as
\begin{align}
    f({\bf x}) &\sim \mathcal{GP}(0, k({\bf x}, {\bf x}' ; {\boldsymbol{\eta}})), \label{eq:gp_prior}
\end{align}
where $f$ is a function drawn from a GP prior with mean zero, $k$ is a covariance function or kernel, and ${\boldsymbol{\eta}}$ are the parameters of that kernel.
This means that the function $f$ is drawn from a probability distribution over a space of functions, where that distribution has a mean of zero and a covariance $k$.
In the case that $f$ is evaluated at a finite set of points $Y_{i} = f({\bf x}_i)$, the distribution over the function values is the multivariate Gaussian distribution $\mathcal{N}({\bf 0}, {\bf K})$, with a covariance matrix $[{\bf K}]_{ij} = k({\bf x}_i, {\bf x}_j)$.

Crucially, as seen in Eqn~\eqref{eq:gp_prior}, GPs specify the distribution of function \emph{by their covariance structure}, which is determined by the choice of kernel function $k$.
This allows us to encode our belief that nearby spectra are more similar, by using GPs to assume that the covariance for each spectrum parameter decreases with distance on the sky.
Kernels that have this property are said to be stationary, meaning that the covariance function depends only on the distance between points.
There are many commonly used stationary kernels where the covariance decreases with distance \citep[see][for a comprehensive overview]{rasmussen2006}.
Here, we will use the Matern-3/2 kernel with the Euclidean distance metric for all GP components
\begin{align}
    k (r_{ij} \,;\, \boldsymbol{\eta}) &= s^2 \left( 1 + \frac{\sqrt{3} r_{ij}}{\ell} \right) \exp{\left( -\frac{\sqrt{3} r_{ij}}{\ell} \right)}, \label{eq:matern} \\ 
    r_{ij} &= ||{\bf x}_i - {\bf x}_j||_2, \label{eq:metric}
\end{align}
which has a lengthscale parameter $\ell$ and a variance parameter $s^2$ (so $\boldsymbol{\eta} = [\ell, s]$).
It is common for parameters of a GP covariance kernel to be called hyperparameters, but we will exclusively refer to them as kernel parameters.
The Matern-3/2 kernel yields functions that are everywhere continuous and once-differentiable.
In practical terms, this means it represents functions that are ``rough'', but not ``spikey'', which is desirable for representing the structure of nebulae, which while having coherent spatial features can be rough or bumpy at small scales.

Since the $f_\mu$ (where we have reintroduced the component index $\mu$, as these choices can differ for each spectrum component $\psi_\mu$) generated via this process may take any value $(-\infty, \infty)$, we include the transformations $T_\mu$ to apply a possible desired constraint to the spectrum parameter function $\psi_\mu$:
\begin{align}
    \psi_{\mu} ({\bf x}) &= T_{\mu}  \left(f_{\mu} ({\bf x})\right). \label{eq:transformed_gp}
\end{align}
For example, if a $\psi_\mu$ represents a physical quantity that is strictly positive (e.g. emission line flux, line width, etc.) then we may pick a transformation $T_\mu: (-\infty, \infty) \to (0, \infty)$.
On the other hand, if a $\psi_\mu$ represents a quantity that can be positive or negative (e.g. line-of-sight velocity), then we may pick $T_\mu$ to be the identity function.
Strictly speaking, each $T_\mu$ will need to be bijective (invertible), monotonic, and everywhere differentiable.

These steps modify the modelled data generation process from Section~\ref{sec:forward_models} in that it is no longer independent for each spaxel.
Instead, the model has shared representations for each spectrum component, drawn from GPs with covariance that depends on the distance between spaxels.
These GPs are \emph{latent}, meaning that they do not model the observed data directly, but rather underlying functions that generate the observed data.

\subsection{Adaptive Spatial Resolution} \label{sec:adaptive_res}

A property of this approach is that the \emph{effective spatial resolution} obtained by the model will naturally adapt based on the S/N of the data.
First, we define the following power spectral densities (PSDs):
\begin{itemize}
    \item \textbf{Noise PSD}: The PSD or Fourier transform of the noise process in the spatial dimensions/sky plane. For our assumed Gaussian white noise, this is a frequency independent constant $N_0$. If the instrument plus atmosphere point spread function (PSF) is significant compared to the fibre spacing, this would instead be a function of spatial frequency determined by the PSF.
    \item \textbf{Kernel PSD}: The PSD of the GP covariance kernel, denoted $\tilde k (\boldsymbol{\omega})$, where $\boldsymbol{\omega}$ is the spatial frequency vector $\boldsymbol{\omega} = [\omega_\alpha, \omega_\delta]$. This encodes our prior belief about the similarity of nearby spectra, and is where the stationarity assumption applies.
    \item \textbf{Effective PSD}: The PSD of the model predictions for each GP component fitting the model, denoted $\tilde h (\boldsymbol{\omega})$. We define this more carefully below, but its Fourier transform $h({\bf x})$ determines the spatial resolution of the predictions and for that reason we refer to it as the effective PSF (ePSF).
\end{itemize}

We can relate each of the above quantities by making a few simplifying assumptions.
First, we assume that $N_0$ is a representative average variance across spaxels and wavelengths, which is reasonable as long as we consider a small wavelength range where the noise properties don't vary too much.
Second, we assume that the model depends linearly on the GP components, which is true if they enter linearly in $g$, and for small perturbations around the mean prediction.
This second assumption is violated in practice for the models we use in our analysis, since all components will either have non-linear transformations applied (positivity for line flux), or enter non-linearly in $g$ (redshifts, line width variations).
However, we can still get some idea of the qualitative behaviour by considering the linear case.

\citet{sollich2004} use linear filter theory to show that the effective PSD (called the ``equivalent kernel'' PSD in their work) of the GP posterior mean prediction is given by
\begin{align}
    \tilde h (\boldsymbol{\omega}) &= \frac{\tilde k (\boldsymbol{\omega})}{\tilde k (\boldsymbol{\omega}) + N_0}, \label{eq:res_scaling}
\end{align}
which demonstrates that the effective PSD is a weighted combination of the kernel and noise, and so the ePSF naturally trades off between S/N and spatial resolution.

In Appendix~\ref{sec:effective_psd} we show that for the Matern-3/2 kernel, the effective lengthscale $\ell_{\mathrm{eff}}$ obtained by the model should scale
\begin{align}
    \ell_{\mathrm{eff}} \propto \mathrm{SNR}^{-2/5}, \label{eq:leff_snr}
\end{align}
where $\mathrm{SNR}$ is the spectrally integrated S/N.
Note that while the kernel could be the same for each GP component, in practice $\ell_{\mathrm{eff}}$ will vary between components due to how well each component is determined by the data.
This will change the proportionality constant in Eq.~\eqref{eq:leff_snr}.
Non-linearities and interactions between components in $g$ (i.e. the redshift estimate is sensitive to the line width estimate) will also modify the scaling from exactly $\mathrm{SNR}^{-2/5}$ to $\mathrm{SNR}^{-q}$ where $q > 2/5$.
Although it's not immediately apparent how non-linear each component is, the line flux is only weakly non-linear (due to the positivity transformation), while redshift and especially line width are probably more non-linear.
We therefore expect that $q$ will be closest to $2/5$ for the line flux component.


\section{Implementation} \label{sec:implementation}

The main challenge for our approach is keeping inference tractable over many spectra, given that naively the GP inference cost scales as $\mathcal{O} (N^3)$, and the memory usage $\mathcal{O} (N^2)$, where $N$ is the number of spaxels.
This is compounded by the fact that each per-spaxel component adds $N$ additional parameters to the model.
Our implementation addresses both of these issues by leveraging state-of-the-art Gaussian process acceleration, and gradient-based optimisation through automatic differentiation and hardware acceleration.

\subsection{Equispaced Fourier Basis}

We use the equispaced Fourier basis representation of a Gaussian process \citep[EFGP;][]{hogg2021,greengard2022,barnett2023}, which rewrites Eq.~\eqref{eq:gp_prior} in an equivalent form
\begin{align}
    f ({\bf x}) &= \sum_{l=1}^{p_\alpha p_\delta} X_{l} \exp{\left(- \boldsymbol{\omega}_l \boldsymbol{\cdot} {\bf x}\right)}, \label{eq:efgp} \\ 
    X_l &\sim \mathcal{N} \left( 0,  \tilde k \left( {\boldsymbol{\omega}}_l ; {\boldsymbol{\eta}} \right)\right),
\end{align}
where ${\tilde k}$ is the PSD of the covariance kernel ${k}$, $X_{l}$ are coefficients for basis function $l$, and $\boldsymbol{\omega}_l \in \Omega_\alpha \times \Omega_\delta$ are spatial frequencies on a square grid in Fourier space
\begin{align}
    \Omega_d = \left\{ -\left(p_d - 1\right) / 2, \dots, \left(p_d - 1\right)  / 2 \right\},
\end{align}
where $p_d$ is the number of modes in direction $d \in \{\alpha, \delta\}$, and $p_d$ is odd.
Eq.~\eqref{eq:efgp} is equivalent to a discrete Fourier transform (DFT).
Specifically, it is the forward non-uniform DFT (NUDFT) in the case where the sky positions ${\bf x}_i$ do not happen to lie on a square grid.
The forward NUDFT is sometimes also known as the ``type-II'' NUDFT, and it maps from uniform grids to non-uniform grids, while the adjoint or ``type-I'' NUDFT does the reverse (although unlike the uniform case, these are not inverses of one another).
The 2D forward NUDFT we use above can be interpreted simply as the evaluation of a Fourier series at arbitrary 2D spatial points, given a set of Fourier coefficients corresponding to a square grid in frequency space.
For LVM, the sky coordinates of the fibre positions for each pointing are arranged on a hexagonal grid with equal spacing between fibre positions \citep{herbst2024}, which motivates the need for evaluating on non-uniform points.
It also means that our method generalises to any arrangement of points in the sky.

Naively the NUDFT requires a very large matrix product, which becomes computationally intractable for large datasets, due to the roughly quadratic scaling in number of points.
We therefore utilise the Flatiron Institute Non-Uniform Fast Fourier Transform (\codesnip{fiNUFFT}) \citep{barnett2019,barnett2020} to evaluate the model efficiently, providing almost linear scaling with respect to the number of spaxels and modes included.
The EFGP + \codesnip{fiNUFFT} setup is best-in-class for inference with Gaussian processes on very large datasets.
It has been shown to scale well even to billions of data points \citep{greengard2022}, and has well-known and controllable approximation error for common kernels \citep{barnett2023}.

In full generality, $X_{l}$ are complex numbers, giving $2 p_\alpha p_\delta$ degrees of freedom for each GP component due to the real and imaginary parts.
However since we are only interested in representing real functions, we can use the conjugate symmetry of the Fourier transform to reduce the number of parameters to just $p_\alpha p_\delta$.
Additionally, one must choose the number of modes $p_\alpha$ and $p_\delta$ to include in the model.
This choice effectively comes down to the choice of lengthscale of the covariance kernel, as smaller lengthscales require higher frequency modes in order to be represented accurately (note also that this representation is only appropriate for stationary kernels).

\subsection{Posterior Objective}

For Bayesian forward models, fitting the model to the data means calculating the posterior distribution over the model parameters given the data using Bayes' theorem, giving
\begin{align}
    p ({\boldsymbol{\xi}, {\boldsymbol{\eta}}} \, | \, {\bf d}) = \frac{p ({\bf d} \, | \, {\boldsymbol{\xi}}, {\boldsymbol{\eta}})  \, p ({\boldsymbol{\eta}})\, p ({\boldsymbol{\xi}})}{p ({\bf d})},
\end{align}
where $\boldsymbol{\xi}$ denotes the non-continuous spectrum components, $\boldsymbol{\eta}$ are the parameters of the GP kernels, and we note the lack of a dependence on the latent GP Fourier coefficients ${\bf X}$.
This is because the typical move for Gaussian process models is to marginalise (integrate out) the latent GPs
\begin{align}
    p ({\bf d} | {\boldsymbol{\eta}}, {\boldsymbol{\xi}}) &= \int p ({\bf d} \,|\, {\bf X}, {\boldsymbol{\eta}}, {\boldsymbol{\xi}}) \, p ({\bf X} \, | \, \boldsymbol{\eta}) \, d {\bf X}, \label{eq:marginal_like}
\end{align}
where $p ({\bf d} \,|\, {\boldsymbol{\eta}}, {\boldsymbol{\xi}})$ is called the marginal likelihood.
The prediction of the GPs themselves (given ${\bf d}, {\boldsymbol{\eta}}, {\boldsymbol{\xi}}$) is then the mean of the Fourier coefficients over the posterior distribution
\begin{align}
    \mathbb{E}({\bf X} \, | \, {\bf d}, {\boldsymbol{\eta}}, {\boldsymbol{\xi}}) = \frac{\int {\bf X} \, p ({\bf d} \,|\, {\bf X}, {\boldsymbol{\eta}}, {\boldsymbol{\xi}}) \, p ({\bf X} \, | \, \boldsymbol{\eta}) \, d {\bf X}}{p ({\bf d} \,|\, {\boldsymbol{\eta}}, {\boldsymbol{\xi}})}, \label{eq:gp_mean}
\end{align}
where the mean ${\bf X}$ values can be interpreted as the expected Fourier coefficients of the GPs given the data and kernel parameters.
The actual predictions for the GP components $\boldsymbol{\psi}({\bf x})$ can then be obtained by feeding these mean values into Eq.~\eqref{eq:efgp} and Eq.~\eqref{eq:transformed_gp}.

If the spectrum model $g$ depended on the GP components only linearly, Eq.~\eqref{eq:marginal_like} and Eq.~\eqref{eq:gp_mean} would have closed form solutions, since all of the probability distributions involved would be Gaussian.
However for our setup this is not the case, since as previously mentioned the spectrum model we will use depends non-linearly on the GP components.
This modifies our likelihood $p ({\bf d} \,|\, {\boldsymbol{X}}, {\boldsymbol{\eta}}, {\boldsymbol{\xi}})$ such that it is no longer Gaussian (the data are still Gaussian distributed, the likelihood function is no longer Gaussian in the model parameters).
We therefore approximate the posterior mean with the maximum a posteriori (MAP) estimate, which corresponds to the mode of the posterior distribution
\begin{align}
    \hat{\boldsymbol{X}} &= \argmax_{{\boldsymbol{X}}} p ({\boldsymbol{X}} \, | \, {\bf d}, {\boldsymbol{\eta}}, {\boldsymbol{\xi}}), \\
    &= \argmax_{{\boldsymbol{X}}} p ({\bf d} \, | \, {\boldsymbol{X}}, {\boldsymbol{\eta}}, {\boldsymbol{\xi}}) \, p ({\boldsymbol{X}} \, | \, {\boldsymbol{\eta}}),
\end{align}
where the second line equality neglects the marginal likelihood normalisation constant (since it does not depend on ${\boldsymbol{X}}$).
In practice, we fix the kernel parameters $\boldsymbol{\eta}$, and jointly optimise the Fourier coefficients $\boldsymbol{X}$ and non-continuous components $\boldsymbol{\xi}$
\begin{align}
    \hat{\boldsymbol{X}}, \hat{\boldsymbol{\xi}} &= \argmax_{{\boldsymbol{X}}, \,{\boldsymbol{\xi}}} p ({\boldsymbol{X}}, {\boldsymbol{\xi}} \, | \, {\bf d}, {\boldsymbol{\eta}}), \\
    &= \argmax_{{\boldsymbol{X}}, \,{\boldsymbol{\xi}}} p ({\bf d} \, | \, {\boldsymbol{X}}, {\boldsymbol{\xi}}, {\boldsymbol{\eta}}) \, p ({\boldsymbol{X}} \, | \, {\boldsymbol{\eta}}) \, p ({\boldsymbol{\xi}}).
\end{align}
The MAP estimates $\hat{\boldsymbol{X}}$ and $\hat{\boldsymbol{\xi}}$ then give our ``best-fitting'' model parameters and determine the predictions.
For computational reasons, in practice we actually \emph{minimise} the negative log posterior
\begin{align}
    \hat{\bf X}, \hat{\boldsymbol{\xi}} &= \argmin_{{\boldsymbol{X}}, \,{\boldsymbol{\xi}}} \left[ \, - {\rm log} \, p ({\bf X}, {\boldsymbol{\xi}} \,|\, {\bf d}, {\boldsymbol{\eta}}) \right]. \label{eq:objective}
\end{align}

The MAP estimate is exactly equal to the mean if the posterior distribution is unimodal and symmetric.
This is \emph{not} the case for our model, again due to the non-linear dependence of the spectrum model on the GP components.
Our predictions should therefore be interpreted as approximating the posterior mean.
We note that this approximation isn't fundamental to the usefulness of the model however, as the MAP estimate is a commonly used point estimate in Bayesian inference in its own right \citep[e.g.][]{gelman2013}.

While this framework gives a natural framework for quantifying uncertainty in the model parameters and predictions, in the form of the posterior distribution, we do not attempt to do so here.
This will involve sampling, at least approximately, from the posterior.
This is a challenging task given the high dimensionality of the EFGP representation, and the non-linear dependence of the model on the latent GP components.
We defer this to a future work, and comment briefly on possible approaches to this problem in Section~\ref{sec:final_steps}.

\subsection{Optimisation} \label{sec:optimisation}

We implement the spectrospatial modelling framework using \codesnip{JAX} \citep{jax2018github}, and the \codesnip{JAX} bindings to \codesnip{fiNUFFT} \citep{barnett2019,barnett2020,jaxfinufft}.
\codesnip{JAX} provides just-in-time (JIT) compilation which dramatically accelerates model evaluations, as well as automatic differentiation allowing for the calculation of gradients of the objective function, and so the use of gradient-based optimisation.
This also allows our models to be easily deployed on hardware accelerators such as graphics processing units (GPUs) and tensor processing units (TPUs), which will be important for applications utilising larger data volumes as well as more complex models.
The framework is built on \codesnip{Equinox} \citep{kidger2021}, which is a general modelling framework for \codesnip{JAX} that allows for easy composition of model ``pieces'' in order to create arbitrarily complicated spectrospatial models, which we further modify to support coupling between parameters/components.
This coupling is used in our analysis to share the kinematics between both lines of the blended \OIIsel doublet (that is, the kinematics of both lines in the doublet point to the same GP components, see Section~\ref{sec:lvm_model}), but could also be useful for future applications directly modelling line ratios as discussed in Section~\ref{sec:discussion}.
Our framework is available as an open-source Python package, \codesnip{spectracles} \citep{hilder2025github}.
The specific code used for the analyses in this paper is also available as a public Zenodo archive \citep{hilder2025a}.

In practice, we also reparameterise the Fourier coefficients $X_l$ to improve numerical stability and convergence during optimisation.
We use the equivalent prior
\begin{align}
    X_l &= \sqrt{\tilde k \left( {\boldsymbol{\omega}}_l ; {\boldsymbol{\eta}} \right)} \, Z_l, \\
    Z_l &\sim \mathcal{N} (0, 1),
\end{align}
and optimise for the values of $Z_l$, which are just rescaled versions of the original Fourier coefficients $X_l$.
This is essentially a form of preconditioning \citep{papaspiliopoulos2007}, and it prevents the gradients from becoming very small for high frequency modes, which would cause very slow convergence.
We note that we still obtain the MAP with respect to the original $X_l$ parameters, since we included the Jacobian adjustment to the objective for this reparameterisation.

We use the library \codesnip{optax} \citep{deepmind2020jax} to optimise the log probability in Equation~\eqref{eq:objective}. 
We use Adam, which is a first order optimiser that adapts the learning rate for each parameter based on estimates of the first and second moments of the gradients \citep{kingma2017}.
Adam is a good choice for our problem since it naturally handles the different scales of the Fourier coefficients and other parameters without tuning, and it scales well to very high dimensional problems.
The framework and implementation are agnostic to the specific choice of optimiser, and the user can easily swap in different optimisers if desired.

Instead of optimising all model parameters simultaneously, we instead use block coordinate descent, which sequentially optimises subsets of the model parameters while keeping the others fixed.
This approach is common in high-dimensional optimisation problems where the model has a very large number of parameters, but where the inherent structure of the model separates those parameters into distinct groups \citep[e.g.][]{bertsekas2016}.
Exploiting this structure can lead to convergence that is both more robust and efficient.
We define the sequence of blocks as the optimisation \emph{schedule}.
For an example model with 2 GP components $\psi_1, \psi_2$, 1 per-spaxel component $\xi_1$, and an additional global (affects all spaxels) parameter $\zeta_0$, the schedule could be:
\begin{align}
    \mathcal{S} = \{ \{ \psi_1 \}, \{ \psi_2 , \zeta_0\}, \{ \xi_1 , \zeta_0\} \}.
\end{align}
The above schedule consists of 3 blocks, each optimising a different subset of parameters.
The first block optimises only the GP component $\psi_1$. 
The second block optimises the GP component $\psi_2$ and the global parameter $\zeta_0$. 
The third block optimises the per-spaxel component $\xi_1$ and the global parameter $\zeta_0$.
This sequence is then repeated enough times such that the objective function and the model parameters converge to desired precision.

Recall that optimising a GP component in our case involves optimising the Fourier coefficients $Z_l$ for that component, where there are $p_\alpha p_\delta$ coefficients per component.
We found that during the first pass through the schedule, we could more robustly achieve convergence by first initialising all $Z_l$ to zero.
Then, once the block where that component is optimised is reached, we set the corresponding $Z_l$ to random values $\sim \mathcal{N} (0, 1)$ and then performed the optimisation.
By more robust convergence, we specifically mean that the optimisation process got stuck in clearly implausible solutions that fit the data poorly less frequently, where these solutions correspond to local minima in the non-convex objective.
The procedure is more often able to avoid bad local minima by allowing each component to find a reasonable solution that well-describes the data, before each additional component adds complexity to the model.

Initialising with random values for the Fourier coefficients also avoids the issue of needing good initialisation, but not wanting to bias the fit towards any ``expected'' solution.
In principle, one could probably replace the schedule approach we describe, which is not very sensitive to initialisation, with one large joint optimisation stage where the components are well initialised based on some expected solution (in LVM, this could be initialising faint lines to have similar structure to bright lines from the same ion).
The schedule approach has the advantage of being able to initialise from random numbers with no assumptions about the expected structure, while only leveraging the structure of the model itself to escape local minima and poor fits.

\section{Application to LVM} \label{sec:methods}

We now apply the framework to build and fit a simple emission line model to LVM data of the Rosette Nebula \citep{drory2024,villa-durango2025}.
Note that our results are not necessarily meant to be full, science-ready data products, but instead a first application of the framework to real data to demonstrate the capabilities of the model.
We model each emission line separately with the spectrospatial framework using a simple Gaussian line profile.
No information is shared between the different emission lines during the fitting process.

\subsection{Line selection} \label{sec:line_selection}

We selected lines to model not according to specific scientific goals, but rather to be representative of the kinds of lines generally of interest in analysis of \HII regions, while also covering a large range of line strengths.
Since the aim of this work is to demonstrate the framework for a very simple model, we will assume each line can be represented by a single Gaussian component.
We therefore avoid any line with multiple components or blends (with one exception, see below and Section~\ref{sec:lvm_model}).
While multiple kinematic components may exist, at the spectral resolution of LVM these are generally unresolved, instead resulting in a single broadened line \citep[e.g.][]{drory2024,kreckel2024}.
For the brighter lines, this is easily confirmed by direct inspection of the spectra.

We selected 23 lines to model, which are listed in Table~\ref{tab:lines}.
These are split into 4 categories in our analysis. 
\begin{enumerate}
    \item \textbf{Strong lines} including the three strongest Balmer lines, and a selection of strong forbidden lines commonly used in nebular analysis. These 9 lines were deliberately chosen to match those shown in the 9-panel LVM overview figure in \citet{drory2024}. For this reason, we also include the \OIIsel doublet. We model this doublet as two lines, but share the kinematics between both lines (and fit them simultaneously, see Section~\ref{sec:lvm_model}).
    \item \textbf{Auroral lines} including \OIIIaur, \NIIaur, \SIIIaur and \OIIaur. These lines are much fainter than the strong lines, but are used for determining the electron temperature in nebulae \citep[e.g.][]{peimbert1967,marino2013,drory2024}.
    \item \textbf{Helium recombination lines} including 6 \ion{He}{1} lines spanning a range of strengths. \HeIi, \HeIii, \HeIiii, \HeIiv and \HeIvi arise from the triplet system, while \HeIv is from the singlet system \citep[e.g.][]{delzanna2022}. \HeIvi is subject to self-absorption effects and deviations from case B recombination \citep[e.g.][]{mendez-delgado2025}. Because of these differing level structures and radiative transfer effects, we do not expect the recovered morphologies and kinematics of all He I lines to be identical, though they should be broadly similar.
    \item \textbf{Metal recombination lines} including \CIIrli, \CIIrlii, \CIIrliii and \NIIrl. These are the faintest of the lines we model. Note that although we will collectively refer to these as metal RLs, \CIIrlii and \CIIrliii are not pure RLs due to significant contributions from fluorescence \citep[e.g.][]{escalante2012,reyes-rodriguez2024}, and so are instead a mix of recombination and permitted components.
\end{enumerate}

\subsection{Data} \label{sec:data}

We used a contiguous 19 pointings of LVM covering the Rosette Nebula, totalling around 35,000 spectra.
These covered the range 3600-9800~\AA{} at a resolution $R\sim4000$, although the resolution varies throughout that range \citep{perruchot2018,drory2024}.
All these spectra are fit simultaneously in our procedure, separately for each line.
Since these data cover multiple pointings taken across multiple nights, in addition to the generic assumptions of the approach outlined in Section~\ref{sec:assumptions}, we will also assume that the underlying astrophysical signal is invariant over the timescale of the observations.

We used the calibrated and sky-subtracted science frames from version 1.1.1 of the LVM data reduction pipeline (DRP; Mejia et al., in prep.).
These data were calibrated and sky-subtracted, although both are still under active development.
We also did not perform any extinction correction to either the spectra or the model, and so all reported results are uncorrected.
Each LVM pointing has a $\sim$0.5$^\circ$ footprint, and the total area covered by the pointings was around 2$^\circ \,\times$~2$^\circ$.
The point spread function (PSF) of $<$3.5 arcseconds is also much smaller than the fibre diameter and separations at 35.3 and 37 arcseconds respectively \citep{herbst2024}.
This justifies our assumption of independent (uncorrelated) noise processes across spaxels.

\begin{table}
    \centering
    \caption{Line centres (in air), wavelength ranges, and flux scaling factors for each line modelled. $^*$Relative to 3726.03 \AA.}
    \begin{tabular}{|c|c|c|c|}
        \hline
        Line & Centre & Range & Scale \\
        & [\AA] & [\AA] & ${\mathrm{[10^{-14} \, erg \, \text{\AA}^{-1} s^{-1} cm^{-2}]}}$ \\
        \hline
            \forbion{O}{2} & 3726.03, & $+11, -8^*$ & 100 \\
            & 3728.82 & & \\
            \ion{He}{1} & 3819.61 & $\pm$ 8 & 1 \\
            \ion{He}{1} & 4026.19 & $\pm$ 8 & 1 \\
            \ion{C}{2} & 4267.0 & $\pm$ 8 & 1 \\
            \Hgamma & 4340.49 & $\pm$ 8 & 30 \\
            \forbion{O}{3} & 4363.21 & $\pm$ 8 & 1 \\
            \ion{He}{1} & 4471.48 & $\pm$ 8 & 5 \\
            \Hbeta & 4861.36 & $\pm$ 8 & 80 \\
            \forbion{O}{3} & 5006.84 & $\pm$ 8 & 200 \\
            \ion{N}{2} & 5679.56 & $\pm$ 8 & 1 \\
            \forbion{N}{2} & 5754.59 & $\pm$ 8 & 1 \\
            \ion{He}{1} & 5876.0 & $\pm$ 8 & 10 \\
            \forbion{S}{3} & 6312.06 & $\pm$ 8 & 1.3 \\
            \Halpha & 6562.85 & $\pm$ 8 & 500 \\
            \ion{C}{2} & 6578.05 & $+2.5, -6$ & 1 \\
            \forbion{N}{2} & 6583.45 & $\pm$ 8 & 100 \\
            \ion{He}{1}  & 6678.15 & $\pm$ 8 & 5 \\
            \forbion{S}{2} & 6716.44 & $\pm$ 8 & 50 \\
            \forbion{S}{2} & 6730.82 & $\pm$ 8 & 30 \\
            \ion{He}{1}  & 7065.19 & $\pm$ 8 & 3 \\
            \ion{C}{2} & 7236.0 & $\pm$ 8 & 1 \\
            \forbion{O}{2} & 7318.92 & $\pm$ 8 & 2 \\
            \forbion{S}{3} & 9531.1 & $\pm$ 8 & 100 \\
        \hline
    \end{tabular}
    \label{tab:lines}
\end{table}

We restricted the data used in each fit spectrally to a small range centred on each line.
We excluded spaxels where the median flux density in the wavelength range is not between $-10^{-14}$ and $5\times10^{-14} \, \mathrm{erg \, \text{\AA}^{-1} \, s^{-1} \, cm^{-2}}$.
The lower bound excluded any spaxels with unrealistically negative fluxes, potentially caused by poor sky subtraction, and the upper bound excluded any spaxels obviously dominated by stellar contributions, since we do not model the stars.
In principle, we could have used and subtracted upstream model spectra that were fit to spaxels containing stars, such as from the LVM data analysis pipeline \citep[DAP;][]{sanchez2024}, but we just excluded the spaxels for simplicity.
We also included a per-spaxel constant offset nuisance parameter, described in the following section, which helps account for stellar continuum in included spaxels.
Any individual pixels within the wavelength range, or entire spaxels, that were masked or flagged by the DRP were also excluded from our fit for that line.

Before fitting, we scaled the fluxes such that the peak brightness of the line is approximately unity, which helped to stabilise and speed-up the optimisation due to the very high dimensionality of the model.
For very weak lines where it is not possible to estimate the peak brightness directly from the data before scaling (e.g. metal recombination lines), we just adopted a fixed scaling of $10^{-14} \, \mathrm{erg \, \text{\AA}^{-1} \, s^{-1} \, cm^{-2}}$.
We also linearly transformed (scale plus shift) the spatial coordinates for all spaxels $(\alpha, \delta)$ to fit within the domain requirements of \codesnip{fiNUFFT}, which requires the points to be contained within the box $[-\pi, \pi] \times [-\pi, \pi]$.
The line centres, wavelength ranges, and flux scaling factors for each line we analysed are given in Table~\ref{tab:lines}.

\subsection{Emission line model} \label{sec:lvm_model}

The parametric model for each line is a Gaussian function with a constant offset nuisance parameter $\Gamma(i)$, given by
\begin{align}
    g  \left(\lambda ; {\boldsymbol{\theta}}_i \right) &= \frac{F ( {\bf x}_i )}{\sqrt{2 \pi \left[\sigma ({\bf x}_i)\right]^2}} \exp{\left( -\frac{1}{2} \left[ \frac{\lambda - \mu ({\bf x}_i)}{\sigma ({\bf x}_i)} \right]^2\right)} + \Gamma(i). \label{eq:lvm_model}
\end{align}
The line fluxes $F ({\bf x}_i)$ are represented with the $\psi_F$ GP component, which is subject to a positivity constraint given by $T^+$
\begin{align}
    F ({\bf x}_i) &= F_{\mathrm{cal}}(j) \, \psi_F \left( {\bf x}_i \right),  \\
    &= F_{\mathrm{cal}}(j) \, T^{+} \left( f_F \left( {\bf x}_i \right) \right),
\end{align}
where $f_F$ is the untransformed GP, we take $T^{+}$ to be the softplus function
\begin{align}
    T^{+} (x) &= \log \left( 1 + \exp \left( x \right) \right),
\end{align}
and $F_{\mathrm{cal}}(j)$ is a per-pointing multiplicative nuisance term that accounts for per-pointing differences in the flux calibration, with $j\in[1,...,19]$ indexing the 19 different pointings.
The $F_{\mathrm{cal}}(j)$ values were inferred simultaneously with the other parameters for a bright line, and then fixed for the other lines, see Appendix~\ref{sec:calibration}.

The observer-frame line centres $\mu ({\bf x}_i)$ are given by total radial velocity $v_{\rm tot} ({\bf x}_i)$
\begin{align}
    \mu ({\bf x}_i) &= \mu_{\rm rest} \left( 1 + \frac{v_{\rm tot} ({\bf x}_i)}{c} \right),
\end{align}
where $\mu_{\rm rest}$ is the rest-frame line centre and $c$ is the speed of light.
The total radial velocity is itself a sum of four components
\begin{align}
    v_{\rm tot} ({\bf x}_i) &= v ({\bf x}_i) + v_{\mathrm{bary}}(i) + v_{\mathrm{cal}}(k) + v_0
\end{align}
where $v ({\bf x}_i)$ is given by an unconstrained GP component
\begin{align}
    v ({\bf x}_i) &= \psi_v \left( {\bf x}_i \right), \\
    &= f_v \left( {\bf x}_i \right),
\end{align}
and represents the intrinsic motion of the nebula.
It can be interpreted as the line-of-sight velocity of the emitting gas at coordinate ${\bf x}$ with respect to the overall mean velocity of the nebula.
$v_{\mathrm{bary}}(i)$ is a per-spaxel barycentric correction, due to the different observation time of each pointing and the wide-field of the IFU within a single pointing.
These values were calculated using \codesnip{AstroPy} \citep{astropycollaboration2013,astropycollaboration2022} for each spaxel, and fixed before fitting the models.
$v_{\mathrm{cal}}(k)$ is a per-spectrograph wavelength calibration nuisance term, of which there are three total.
$v_0$ is an global systematic velocity term shared across all spaxels, which accounts for the bulk motion of the nebula as seen by the solar system barycentre.

The per-spectrograph wavelength calibration parameters $v_{\mathrm{cal}}(k)$ account for systematic differences in the wavelength calibration in the blue arm across the three spectrographs (a known issue in the current DRP version), where $k\in[1,2,3]$ indexes the spectrographs.
These parameters are inferred simultaneously with the other parameters for a bright line, and then fixed for the other lines, see Appendix~\ref{sec:calibration}.

The observer-frame line widths are given by the measured line spread function (LSF) width $\sigma_{\rm LSF}$, as calculated by the LVM DRP, summed in quadrature with a strictly positive broadening component due to the intrinsic velocity dispersion of the nebula $\sigma_v ({\rm x}_i)$
\begin{align}
    \left[ \sigma ({\bf x}_i) \right]^2 &= \left[ \sigma_{\rm LSF} \left( {\bf x}_i \right) \right]^2 + \left[ \mu({\bf x}_i) \frac{v_\sigma \left( {\bf x}_i \right)}{c}  \right]^2,
\end{align}
where $v_\sigma \left( {\bf x}_i \right)$ is given by a positively constrained GP component for the intrinsic velocity dispersion
\begin{align}
    v_\sigma \left( {\bf x}_i \right) &=  \psi_{v\sigma} \left( {\bf x}_i \right), \\
    &= T^{+} \left(f_{v\sigma} \left( {\bf x}_i \right)\right),
\end{align}
where $T^{+}$ is again the softplus function, and $f_{v\sigma}$ is the unconstrained GP.
This means that the inferred dispersion $v_\sigma \left( {\bf x} \right)$ can be interpreted as the line-of-sight velocity dispersion of the emitting gas at sky coordinate ${\bf x}$, after accounting for the instrumental broadening.

The spatially continuous components are therefore $\boldsymbol{\psi} = \left[ \psi_F, \psi_v, \psi_{v\sigma} \right]$, and the per-spaxel and global components are $\boldsymbol{\xi} = \left[ \Gamma, v_0 \right]$.
We do not include the per-spaxel barycentric corrections, the per-pointing flux calibration corrections, or the per-spectrograph wavelength calibration corrections in $\boldsymbol{\xi}$ since they are fixed during optimisation (see Appendix~\ref{sec:calibration}, for the calibration factors).

The exception to this setup is for the blended \OIIsel doublet, where the flux for each line is modelled separately, but the kinematics are assumed to be identical.
Eq.~\eqref{eq:lvm_model} is modified to include two Gaussians (one for each line), while keeping the single set of offsets.
Both of these components were still fit simultaneously to the data within the wavelength range given in Table~\ref{tab:lines}.
The spatially continuous components for the doublet are therefore $\boldsymbol{\psi}_{\rm doublet} = \left[ \psi_{F1}, \psi_{F2}, \psi_v, \psi_{v\sigma} \right]$, while $\boldsymbol{\xi}$ is unchanged.

We used the same shared lengthscale $\ell$ for the covariance kernel across all components, set to 124 arcseconds, or about 3.4 times the LVM IFU fibre spacing of 37 arcseconds \citep{herbst2024}.
This corresponds to a prior belief that the spatial structure of the emission lines is similar across a distance of around a few spaxels.
We set the kernel standard deviations to 1 in scaled flux units for the flux, 1 $\mathrm{km \, s^{-1}}$ for the radial velocities, and to 3 $\mathrm{km \, s^{-1}}$ for the velocity dispersions component.
This corresponds to a prior belief that the flux varies on the order of the peak flux, the radial velocities vary on the order of a few $\mathrm{km \, s^{-1}}$, and the velocity dispersions vary on the order of a few $\mathrm{km \, s^{-1}}$.
See Section~\ref{sec:discussion} for further discussion of these choices.

The global systematic velocity $v_0$ was left free and separately inferred for each line.

For all GP components we used a grid of $p_\alpha = p_\delta = 401$ Fourier modes in each direction, giving a total of $p_\alpha p_\delta = 160,801$ Fourier coefficients per component.
We verified that this was sufficient for all lines by checking that the highest frequency Fourier coefficients $X_{l}$ converged to zero during the optimisation.
Combined with the per-spaxel offsets, this gives a total of approximately 500,000 model parameters per emission line model.

We also need to specify priors for the per-spaxel offsets $\Gamma(i)$, and the global velocity $v_0$.
For these, we just assumed unbounded, uniform priors over the real line
\begin{align}
    v_0 &\sim \mathcal{U}(-\infty, \infty), \\
    \Gamma(i) &\sim \mathcal{U}(-\infty, \infty).
\end{align}
We found these to be sufficient in practice as the data strongly constrain these parameters.

Taking together all of the above, and following Equation~\eqref{eq:objective}, the objective function we minimise for each line is
\begin{widetext}
    \begin{align}
        - {\rm log} \, p ({\bf X}, {\boldsymbol{\xi}} \,|\, {\bf d}, {\boldsymbol{\eta}}) &= - {\rm log} \, p ({\bf d} \, | \, {\bf X}, {\boldsymbol{\xi}}, {\boldsymbol{\eta}}) - {\rm log} \, p ({\bf X} \, | \, {\boldsymbol{\eta}}) - {\rm log} \, p ({\boldsymbol{\xi}}), \\
        &= \frac{1}{2} \left[ \sum_{i=1}^N \sum_{j=1}^M \left( \frac{d_{ij} - g(\lambda_j; {\boldsymbol{\theta}}_i)}{\sigma_{ij}} \right)^2 + \log{2\pi \sigma_{ij}^2} \right] + \frac{1}{2} \sum_{\mu} \sum_{l=1}^{p_\alpha p_\delta} \left[ Z_{\mu l}^2 + \log (2 \pi)  \right] + \rm{const.}, \label{eq:lvm_objective}
    \end{align}
\end{widetext}
where $i$ indexes the $N$ spaxels, $j$ indexes the $M$ wavelength pixels in the spectral range around the line, $\mu$ indexes the GP components, and $l$ indexes the Fourier coefficients for each component.
The first term is the negative log likelihood, the second term is the negative log prior for the Fourier coefficients, and the third term is the negative log prior for the per-spaxel offsets and global velocity, which is just a constant here due to the uniform priors.

\subsection{Optimisation schedule}

To fit the emission line models we followed the procedure outlined in Section~\ref{sec:optimisation}, with the following schedule
\begin{align}
    \mathcal{S} = \{ \{ \psi_F, v_0, \Gamma \}, \{ \psi_v \}, \{ \psi_{v\sigma}\} \},
\end{align}
adopting a learning rate of $10^{-2}$ \citep[the global learning rate; the per-parameter learning rate is tuned automatically by Adam, see][]{kingma2017} 2000 steps per block, and a total of two passes through the schedule.
We found that this learning rate and number of steps was sufficient to achieve convergence within each block for all lines, in that both the objective function and the model parameters did not change with further iterations or passes.

For the doublet model, this schedule was modified to
\begin{align}
    \mathcal{S}_{\rm doublet} = \{ \{ \psi_{F1}, v_0, \Gamma \}, \{ \psi_{F2}, v_0, \Gamma \}, \{ \psi_v \}, \{ \psi_{v\sigma}\} \},
\end{align}
with the same learning rate and number of steps per block, and again a total of two passes through the schedule.

The optimisation was performed on a M1 Macbook Pro laptop with 16 GB of RAM, and took around 10 minutes per line, or around 20 minutes for the doublet.

\section{Results} \label{sec:results}
We now present spatially continuous maps of fluxes and kinematics for all lines listed in Table~\ref{tab:lines}.
The maps were computed directly from the model components: the flux intensity maps are calculated with $F({\bf x})$, the radial velocity maps with $v({\bf x})$, and the velocity dispersion maps with $v_\sigma({\bf x})$.
The flux intensity map for the doublet was calculated as $F_1({\bf x}) + F_2({\bf x})$.
We emphasise that these maps were computed \emph{directly} on a dense grid in ${\bf x} = [\alpha, \delta]$ spanned by the LVM pointings, and not interpolated from the fibre positions.

\subsection{Strong emission lines}

\begin{figure*}
    \centering
    \includegraphics[width=0.98\textwidth]{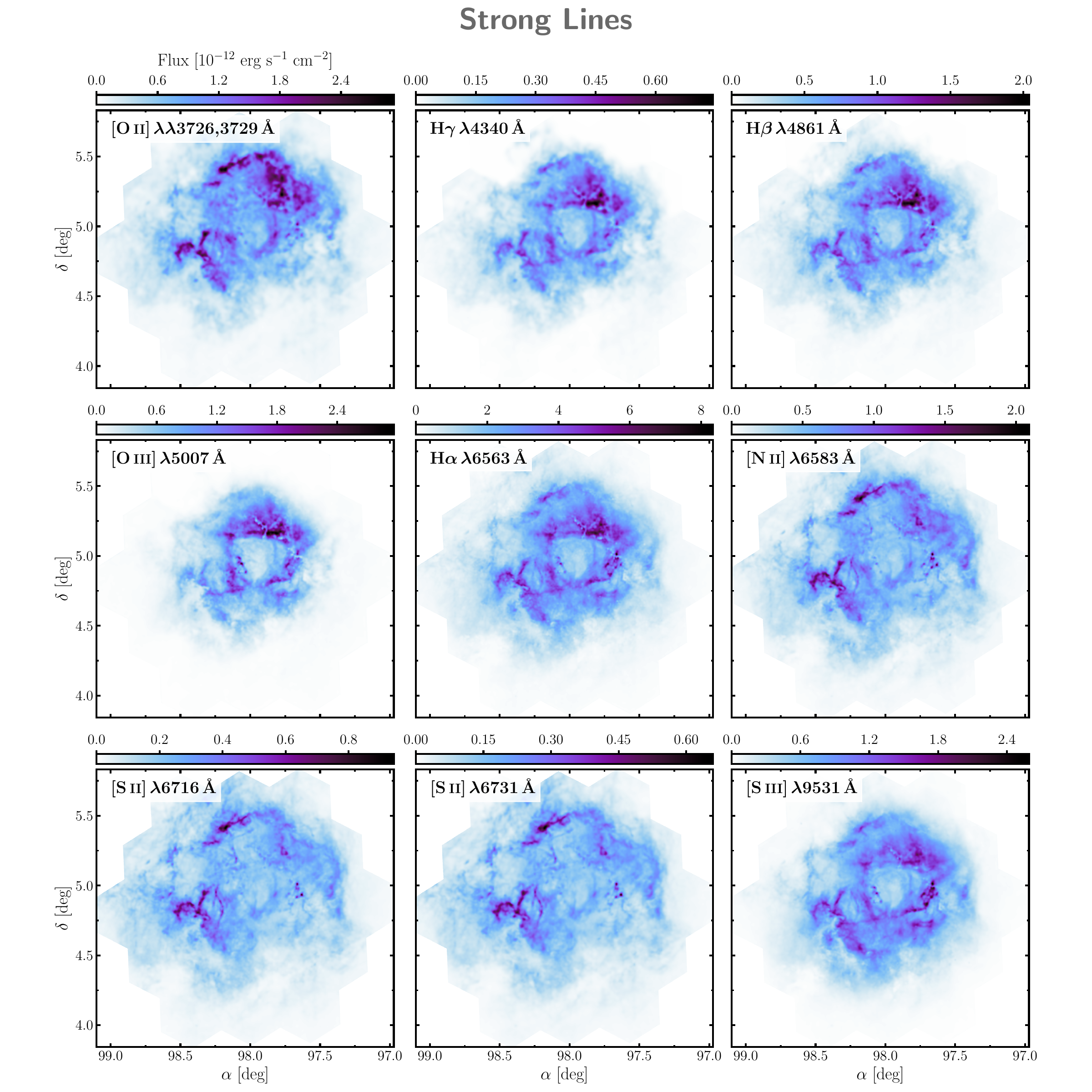}
    \caption{Flux intensity maps for the chosen strong emission lines, in units of $10^{-12} \, \mathrm{erg \, s^{-1} \, cm^{-2}}$. The lines are ordered by increasing wavelength from left to right, top to bottom. The maps cover a contiguous region of the Rosette Nebula, from 19 LVM pointings or ``tiles''. These maps show the flux continuously across the sky, rather than just at the fibre positions. No interpolation is performed; the values are computed directly from the model $F({\bf x})$ inferred from the data, on a dense grid of sky locations. The strength of the lines varies by about one order of magnitude, with \Halpha being the strongest, and \Hgamma the weakest.}
    \label{fig:flux_9lines}
\end{figure*}

Figure~\ref{fig:flux_9lines} shows the flux intensity maps for the nine strong lines presented in the LVM science overview paper \citet{drory2024}.
These maps reveal the expected morphology of the Rosette Nebula to high spatial fidelity, and demonstrate the continuous property of the model predictions.
We do not see any artefacts from missing or excluded data, nor from the joins between adjacent pointings.

The sharpness of these maps is linked to the S/N, of the particular line, which is evidenced by the brighter lines appearing sharper.
The brightest line is \Halpha, where we see the highest effective resolution spatially, even in the very faint emission in the outer regions of the nebula.
\Hgamma is the faintest line and appears somewhat blurrier, although the same structures are still clearly resolved and very fine structures are still visible throughout the nebula.
A quantitative analysis of the effective spatial resolution as a function of S/N is presented in Section~\ref{sec:resolution_results}.

\subsection{Auroral lines}

\begin{figure*}
    \centering
    \includegraphics[width=0.98\textwidth]{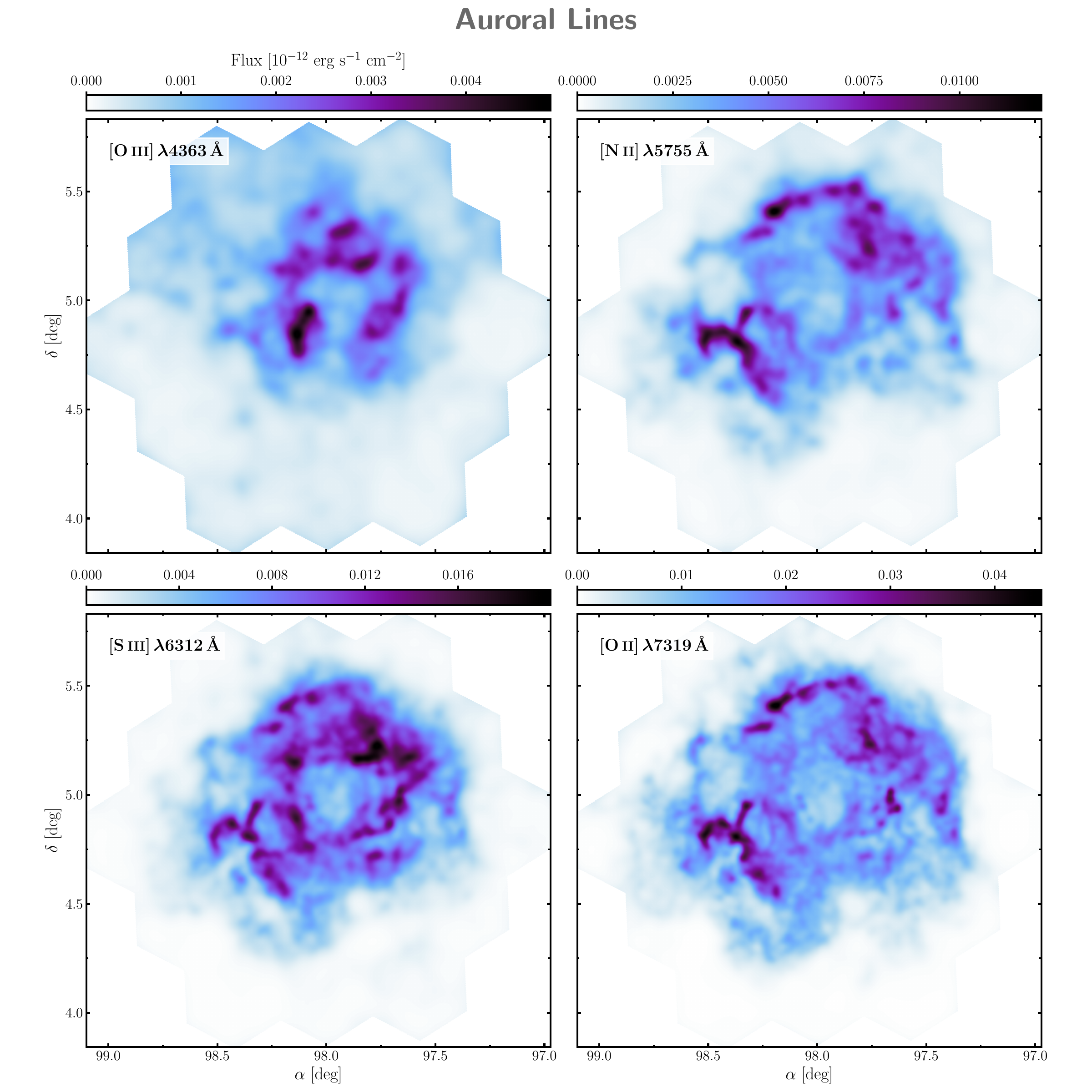}
    \caption{Similar to Figure~\ref{fig:flux_9lines}, but for the auroral lines \OIIIaur, \NIIaur, \SIIIaur, and \OIIaur. The weakest of these is \OIIIaur, which is around 1000 times weaker than \Halpha.}
    \label{fig:flux_auroral_lines}
\end{figure*}

The flux intensity maps for four auroral lines are shown in Figure~\ref{fig:flux_auroral_lines}.
These lines are around 200-2000 times fainter than \Halpha, and we can clearly see that the model trades off the lower S/N of these lines for a lower effective spatial resolution.
Despite this, the spatial structure is still well-resolved, and we see many of the same structures as in the brighter lines. \OIIIaur is the faintest of these lines, and is one of the weakest lines included in our analysis, but we still see that the model has resolved spatial structure in the brighter central region of the nebula, and these structures appear similar to those seen in brighter lines but less well-resolved.

\subsection{Helium recombination lines}

\begin{figure*}
    \centering
    \includegraphics[width=0.98\textwidth]{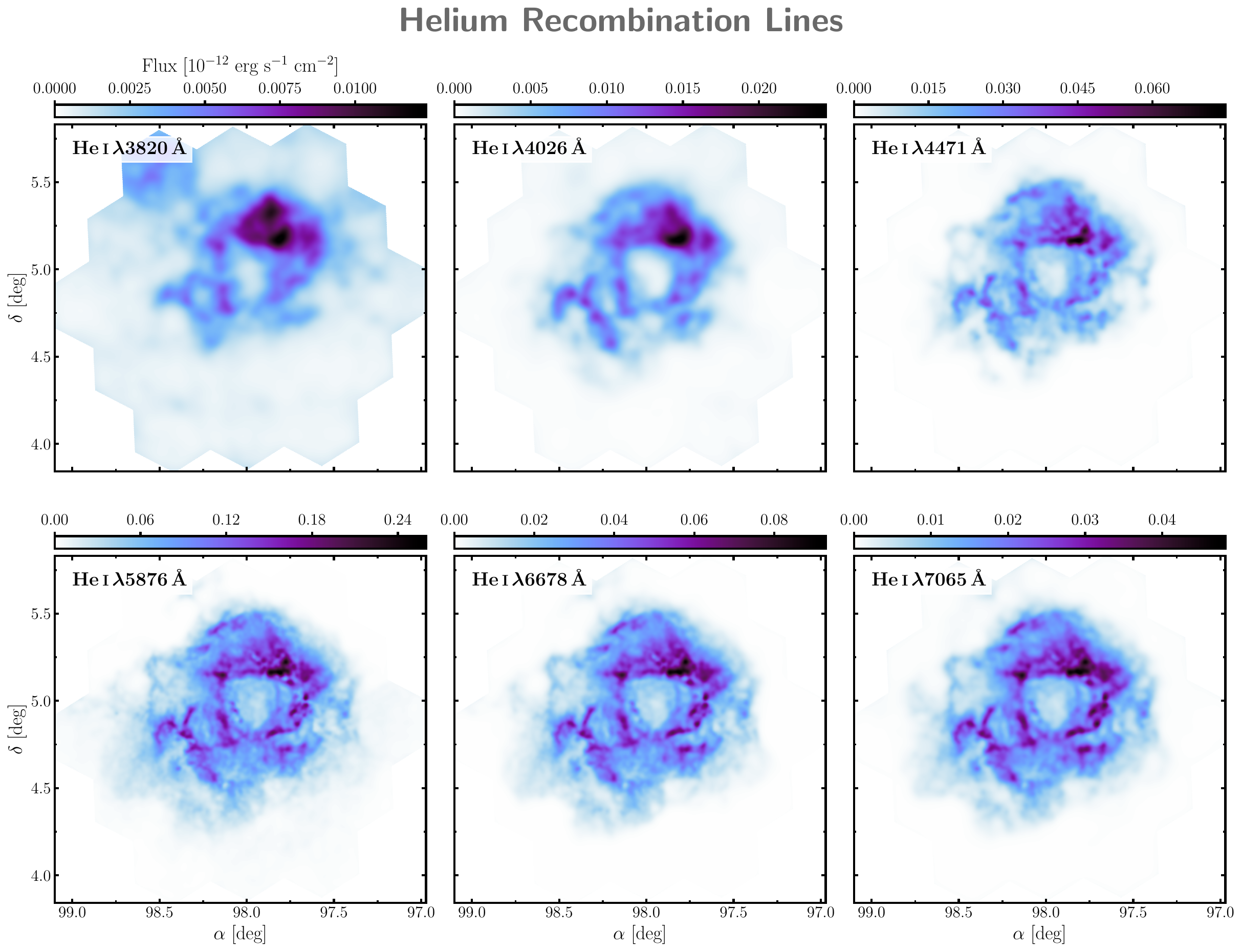}
    \caption{Similar to Figure~\ref{fig:flux_9lines}, but for six \ion{He}{1} recombination lines.}
    \label{fig:flux_helium_lines}
\end{figure*}

Figure~\ref{fig:flux_helium_lines} shows flux maps for six \ion{He}{1} lines at different wavelengths and of varying strengths between $\sim$30 (\HeIiv) and $\sim$800 (\HeIi) times weaker than \Halpha.
These lines arise from the same ionic species (although different level structures, see Section~\ref{sec:line_selection}), and so the underlying spatial and velocity structure should be similar, up to differences in the radiative transfer effects and level populations.
This makes them a great case study for our approach in the medium to low-S/N regime, as the main difference between each line is the S/N.
We see a clear progression in the level of spatial fidelity resolved by each of the maps, with the resolution increasing with the strength of the line.

The bottom row of the figure shows the three strongest HeI lines, where we can see faint structures in the outer regions resolved to a level of detail comparable to \Hgamma.
The top row shows the three weakest lines, where the diffuse emission in the outer regions becomes more difficult to resolve and we only map variations across the central brighter region of the nebula.
The weakest line \HeIi also shows evidence of an unaccounted for systematic effect contaminating the North-East most pointing, as there is an artefact in the form of a systematically brighter hexagonal region covering the footprint of the IFU.

\subsection{Metal recombination lines} \label{sec:metal_lines}

\begin{figure*}
    \centering
    \includegraphics[width=0.98\textwidth]{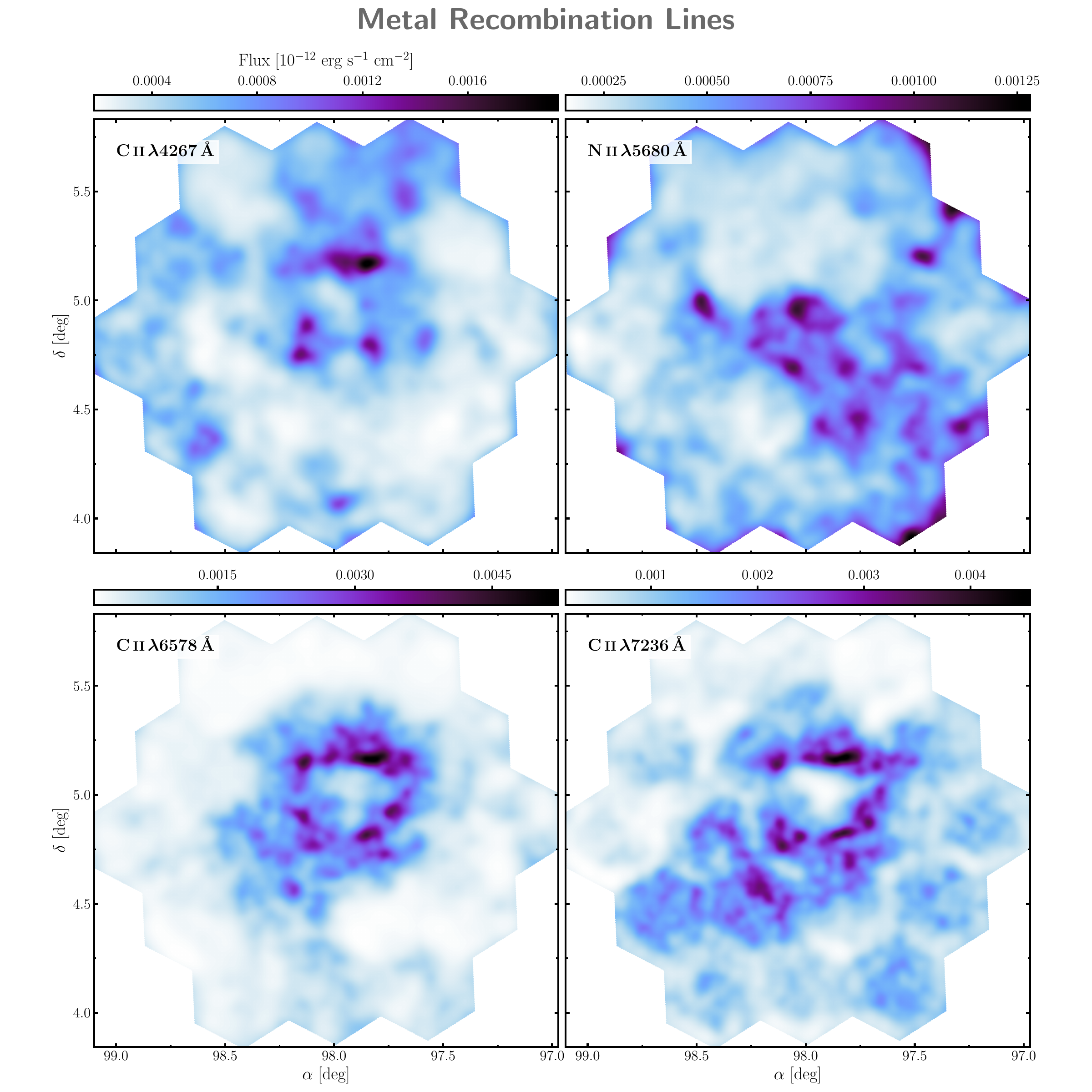}
    \caption{Similar to Figure~\ref{fig:flux_9lines}, but for \CIIrli, \NIIrl, \CIIrlii and \CIIrliii. Note that \CIIrlii and \CIIrliii have both recombination and permitted contributions due to fluorescence. All these lines are at least 1000 times weaker than \Halpha. These maps suffer from some artefacts due to their very low S/N, but the spatial structure of the nebula is still resolved.}
    \label{fig:flux_metal_lines}
\end{figure*}
We present flux intensity maps for four metal recombination lines in Figure~\ref{fig:flux_metal_lines}, where we note again that \CIIrlii and \CIIrliii are not pure recombination lines due to significant contributions from fluorescence (see Section~\ref{sec:line_selection}).
\CIIrlii and \CIIrliii are the strongest lines, with peak fluxes in the map approximately 2000 times weaker than \Halpha, while \CIIrli and \NIIrl are around 2--4 times weaker again.
For both \CIIrlii and \CIIrliii we can clearly see the inner ionised structure of the nebula, as in the other maps of brighter lines.
We emphasise that no information is shared between these fits and so this morphology is found completely independently by the model for each line separately.
The \CIIrlii map appears to be the highest quality.
The \CIIrliii map is similar in strength and morphology, but suffers from some unaccounted for systematic as can be seen by the higher-flux hexagons repeated in the centre of each pointing in the South-West corner of the map.
This is likely caused by sky contamination, either from the subtraction process in the DRP or from the sky lines themselves.

The \CIIrli map also resolves some of the inner ring-like structure of the nebula, but similarly to the \OIIIaur map, it has some systematically brighter pointings around the edge of the maps that are likely artefacts.
Additionally, several of the fibres nearby this line in wavelength have known defects and so are masked by the DRP.
These data are also excluded from our fit, and so it is possible that having more excluded data in this case could be affecting the morphology of the recovered map.

The \NIIrl map is the weakest line included in our analysis, and while the model has found some spatial structure, it does not look similar to any of the other lines.
There are also three dimmer pointings in the central region of the map that are hexagonal shaped and so probably affected by some systematic.

\subsection{Kinematics}

\begin{figure*}
    \centering
    \includegraphics[width=0.98\textwidth]{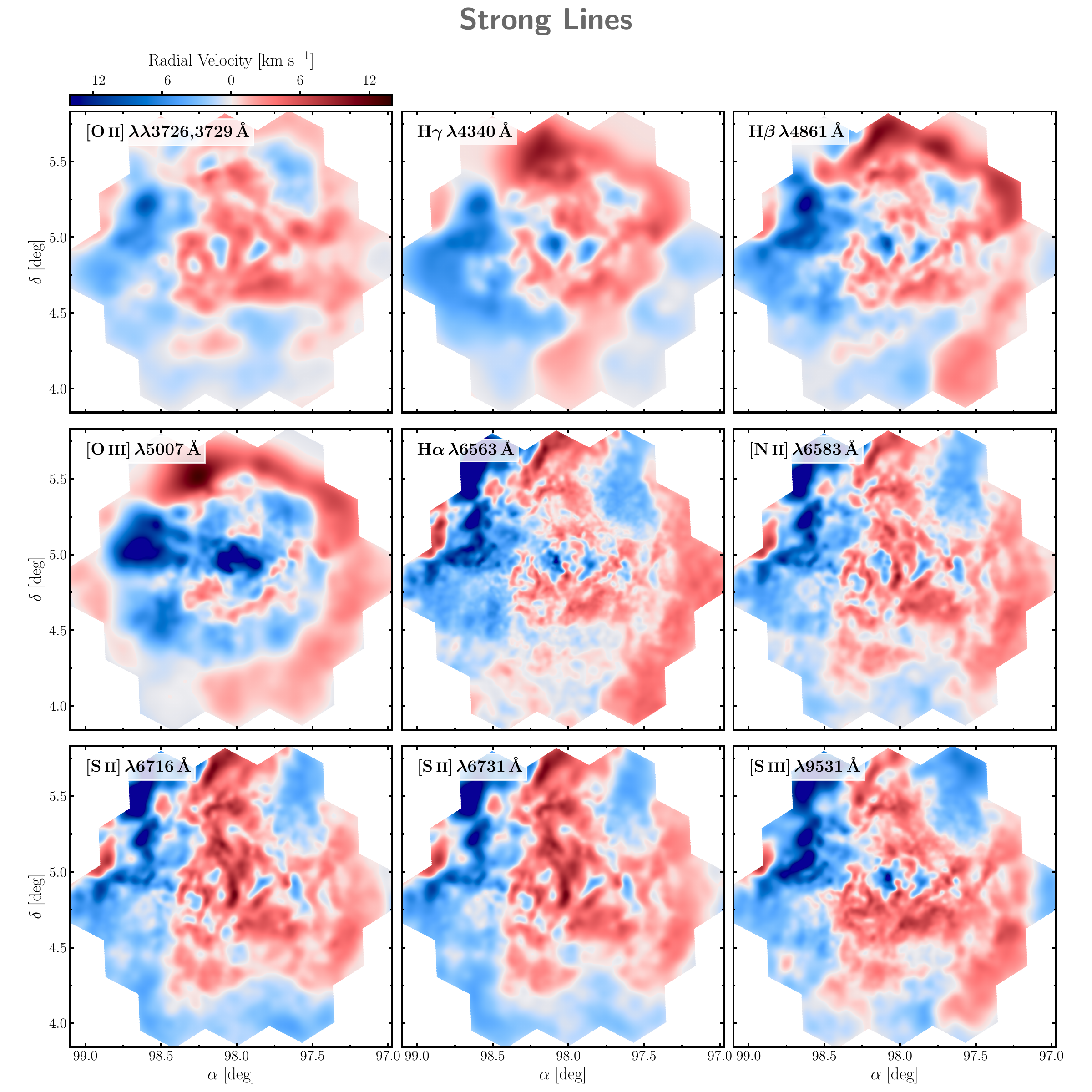}
    \caption{Radial velocity (line-of-sight velocity) maps for the chosen strong emission lines, in units of $\mathrm{km \, s^{-1}}$. These velocities are calculated directly with the best-fit model $v({\bf x})$, and represent the line-of-sight velocity of the emitting gas with respect to the bulk motion of the nebula.}
    \label{fig:vrad_9lines}
\end{figure*}

\begin{figure*}
    \centering
    \includegraphics[width=0.98\textwidth]{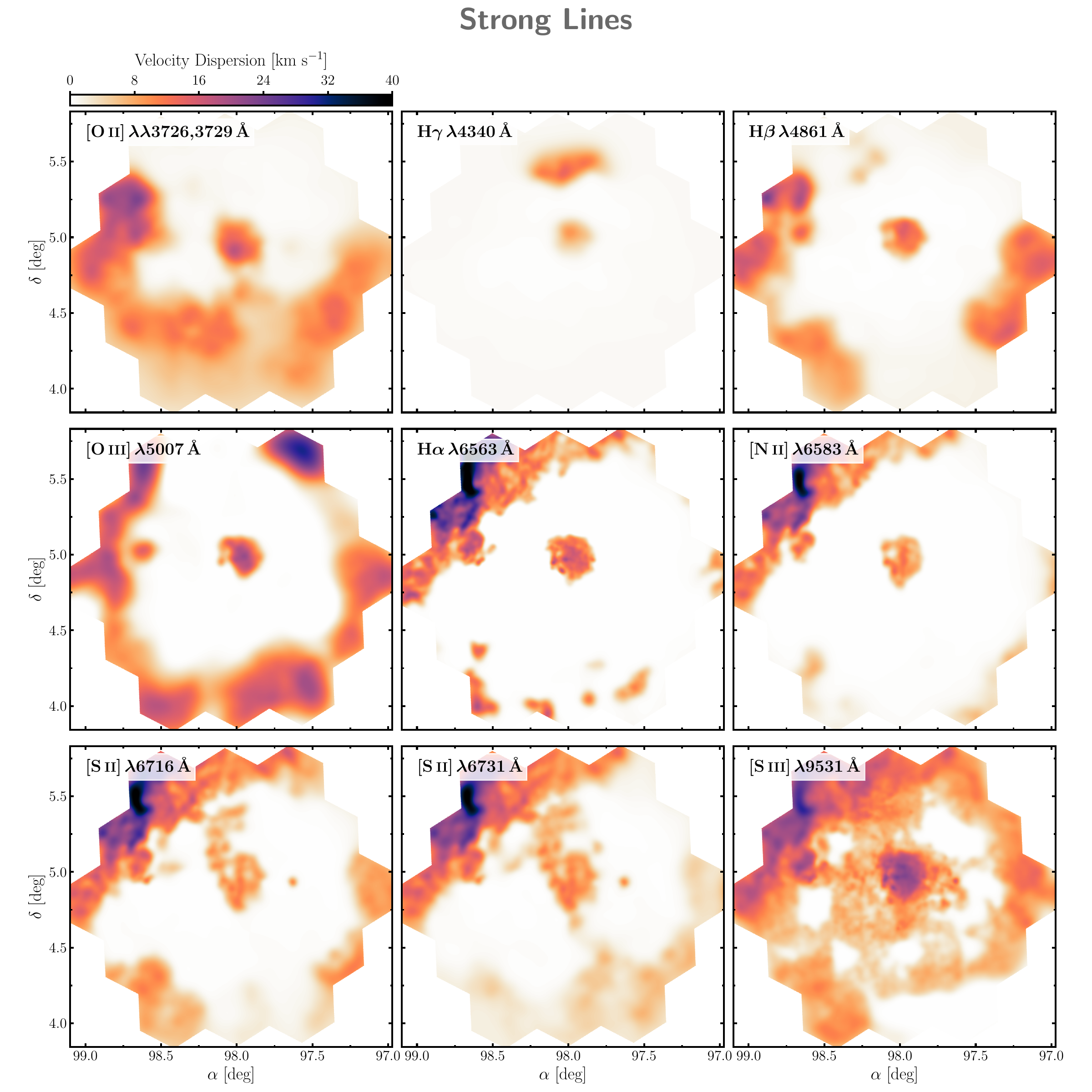}
    \caption{1$\sigma$ velocity dispersion maps for the chosen strong emission lines, in units of $\mathrm{km \, s^{-1}}$. These dispersion maps are calculated directly with the best-fit model $v_\sigma({\bf x})$, and represent the intrinsic line-of-sight velocity dispersion of the emitting gas, accounting for the instrumental broadening.}
    \label{fig:disp_9lines}
\end{figure*}

We present radial velocity and velocity dispersion maps for all nine of the strong lines in Figures~\ref{fig:vrad_9lines} and \ref{fig:disp_9lines} respectively.
We recover the complex radial velocity structure in \Halpha as seen in the science overview paper \citet{drory2024}, with nested shells of rapid sign changes in the central regions and larger flows towards the outer regions of the nebula.
The other Balmer lines, \Hgamma and \Hbeta, show a similar set of structures except at a lower effective spatial resolution.
This preference for weaker velocity variations is due to the implicity regularisation encoded by the Gaussian process priors, which encourage the model towards smoother and closer to zero velocity fields when less informed by the data.

We note as pointed out in Section~\ref{sec:adaptive_res}, there is nothing preventing the model from recovering different spatial resolutions across the different components of the same line, and indeed it is expected that the velocity and dispersion maps will yield lower spatial resolutions than the flux maps for the same lines.
This is clearly born out in the results, as for example the velocity map for \Hbeta appears less structured and more diffuse compared to the flux map.

The difference in resolution between the flux and velocity maps varies across different lines.
For example, the flux and velocity maps for \SIIIsel both appear very well-resolved.
This is likely demonstrating the predicted more-rapid decrease in resolution as S/N decreases for the velocities, which we investigate further in Section~\ref{sec:resolution_results}.

The velocity structures for the lines emitted primarily in the outer regions, \NIIsel, \SIIseli and \SIIselii, of the nebula additionally show strong agreement with each other down to small scales.
The \OIIIsel line is the most different of all the radial velocity maps, with a very blue-shifted central region instead of the varying-sign flows seen in the other lines.
This is surrounded by a large less blue-shifted region, and then a red-shifted outer region.
These bulk flows are not seen as much in the other velocity maps, where the structure is more complex with large variations in velocity on small physical scales.

As with the other maps, the velocity dispersion maps show similar structure across the lines resolved at different effective spatial resolutions.
They are generally smoother and less well-resolved than the flux maps and to a lesser extent the radial velocity maps.
However, the maps are much sparser in terms of detected features, with large portions of the maps set to zero.
This ``floor'' effect also changes with the wavelength of the line, with the maps at larger wavelengths containing larger regions traced at smaller dispersions, seen especially in \SIIIsel.
This is likely driven by the scatter in the values used for the instrumental LSF from the DRP, which is greatest at shorter wavelengths.
This is discussed in detail in Appendix~\ref{sec:lsf}.

The maps contain multiple consistent features across all lines, include a large bubble-like feature occupying the centre-most region of the nebula, with smaller blobs nearby.
The \Halpha map shows some difference to the map in the science overview paper, which contained a large non-zero dispersion region in the South-West region at around $(97.5^\circ, 4.5^\circ)$, whereas our map is largely empty there.
The cause of this is not immediately clear, but it could be due to the scatter in the assumed LSF as already mentioned.
We note that the region \emph{is} seen in the \SIIselii and \SIIIsel dispersion maps.

\subsection{Spatial Resolution} \label{sec:resolution_results}

\begin{figure*}
    \centering
    \includegraphics[width=0.98\textwidth]{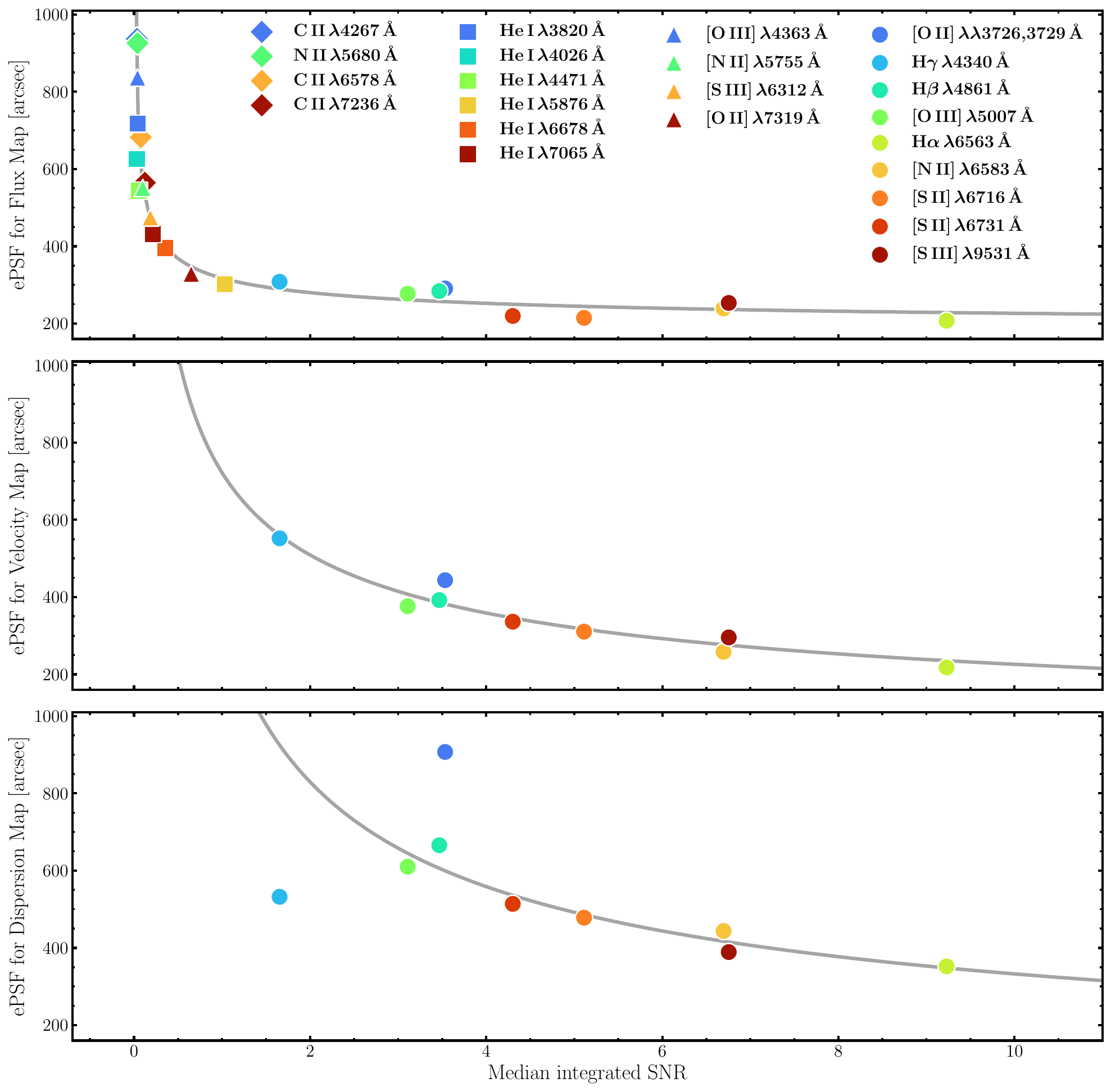}
    \caption{Effective PSF lengthscales for the GP components for each modelled line, as a function of the median spectrally integrated S/N. The top, middle and bottom panels show the results for the flux intensity, radial velocity, and velocity dispersion components respectively. The circle, triangle, square, and diamond markers correspond to the strong, auroral, Helium recombination, and metal recombination lines respectively. Only the strong lines are shown for the velocity and dispersion components. The grey lines show least-squares fits to the points assuming a scaling of the form $\ell_{\rm eff} \propto {\rm SNR}^{-q}$, with the points appearing tightly clustered around the trend lines.}
    \label{fig:epsf}
\end{figure*}

We investigated the relationship between our obtained spatial resolutions and the S/N of the lines by estimating $\ell_{\mathrm{eff}}$ for the GP components of each line.
The method we used to determine these estimates is outlined in Appendix~\ref{sec:effective_psd}, but it is essentially a maximum likelihood estimate assuming that the effective PSD has the same functional form as the kernel PSD.
These estimates are shown as a function of the median spectrally integrated signal-to-noise ratio ${\rm SNR}$ in Figure~\ref{fig:epsf}.
$\ell_{\rm eff}$ is shown for the flux GP component for all lines in the top panel.
For the velocity and dispersions in the middle and bottom panels, we show the resolutions only for strong lines, as we were unable to obtain reliable estimates for the weaker lines due to the very low number of non-zero Fourier coefficients (the velocity and dispersion components are effectively zero for the weaker lines anyway).

Also shown in grey are least-squares trend lines, assuming a scaling of the form we outlined in Section~\ref{sec:adaptive_res} and derived in Appendix~\ref{sec:effective_psd},
\begin{align}
    \ell_{\rm eff} = a \, {\rm SNR}^{-q} + b, \label{eq:res_scaling_fit}
\end{align}
where we fit for $a$, $b$, and $q$ using the values in the respective panels.
We also excluded the \OIIsel and \Hgamma lines from the fit for the dispersions, as they appear as clear outliers.
It is likely that this scatter is driven by quality of the $\ell_{\rm eff}$ determinations, rather than anything about the models themselves.

The resolution scaling in the Figure appears well-described by Eq.~\eqref{eq:res_scaling_fit}, with the points clustered tightly around the trend lines.
The $q$ values obtained for the flux, velocity and dispersion components were $0.44$, $0.50$, and $0.58$ respectively.
This is consistent with the expectations outlined in Section~\ref{sec:adaptive_res}, with the flux scaling remarkably only slightly worse than the predicted $q=2/5$ from linear theory.
The velocity and dispersion scalings are both steeper, with the dispersion resolution degrading the fastest as the S/N decreases.

\section{Discussion} \label{sec:discussion}

\subsection{Spectrospatial emission line maps}

We presented a new approach for modelling IFU data that treats the properties of the spectra as drawn from continuous functions of sky position, where these functions are inferred from the data.
We demonstrated this approach with observations of both strong and weak emission lines in the Rosette Nebula taken by LVM.
The recovered morphology and velocity structure of the nebula is consistent across different lines, even when the S/N varies by multiple orders of magnitude.
For the bright lines, we resolve very fine structures even in the fainter outer regions of the Rosette, and the maps are free from any artefacts caused by the joins between pointings, missing data, or the fibre positions themselves.

As seen for the SELs, the spatial-similarity assumption does not lead to washing out or smoothing over real features in the case that small scale fibre-to-fibre variations are well above the noise level.
That is, the maps are \emph{not just} smoothed versions of per-spaxel approaches, but instead perform reconstruction of the true underlying fields in a robust way.

Although no information is shared between fits to different lines, the model recovers very similar morphologies and velocity structures for lines emitted from the same ionic species, even when the S/N varies by orders of magnitude. 
Lines from different species but arising in similar volumes like [\ion{N}{2}] and [\ion{S}{2}] also show remarkably consistent morphologies even for very fine and faint structures.
This consistency, in combination with the random initialisation, provides additional validation of the technique and assumptions.

We presented flux intensity maps for four metal recombination lines, which are by far the faintest lines in our analysis.
While our results are not perfect and suffer from artefacts that likely arise from small per-pointing systematics, especially in the \CIIrliii map, we still see that the model has resolved spatial structure in the brighter central region of the nebula.
The most robust map is \CIIrlii, where the morphology appears similar to brighter lines at lower spatial resolution, and the map is free from obvious artefacts. 
Unfortunately, the line is not commonly used for abundance diagnostics due to the previously mentioned fluorescence contributions \citep[e.g.][]{escalante2012,reyes-rodriguez2024}.
The other lines we present are standard diagnostics \citep{tsamis2004} and so these maps will be directly useful for probing the cause of the abundance discrepancy problem in \HII regions.

The model could likely be modified to better account for sky-related systematics, which would improve the \CIIrliii map, but this is beyond the scope of the present work. 
These maps will also improve further as the LVM DRP and sky subtraction are refined.
We note that although \CIIrli is expected to be the brightest optical line for an \HII region \citep[e.g.][]{liu2004}, it is the faintest of the three \ion{C}{2} lines we present here.
This is probably due to dust extinction, which is much stronger at bluer wavelengths, since our results are not corrected for extinction.

High precision kinematics are one of the main science goals of LVM, targeting $3\sigma$ precision of $5\,\mathrm{km\,s^{-1}}$ for radial velocities and $12\,\mathrm{km\,s^{-1}}$ for dispersions for \Halpha based on DAP simulations \citep{drory2024,sanchez2024}. 
While we do not provide quantitative precisions here, our velocity and dispersion maps show many features varying on scales much smaller than these values, and these features are consistent across different lines. 
This strongly indicates that the approach is providing very high precision kinematics even for lines weaker than \Halpha like \NIIsel, \SIIseli and \SIIIsel, though dispersion precision remains limited by scatter in the instrumental LSF (see Appendix~\ref{sec:lsf}).

\subsection{Resolution-information trade-off}

For the auroral lines in our analysis, the brightest of which are still an order of magnitude fainter than the faintest of the strong lines, we recover essentially the same morphology as for stronger lines of the same species, but at a lower effective spatial resolution.
This is a key feature of our approach: as the S/N decreases, the model compensates by combining information across larger spatial scales.
This is demonstrated visually by the flux maps for the \ion{He}{1} RLs, where identical morphology is recovered at different resolutions depending on the S/N of the line, and quantitatively by the effective resolution estimates shown in Figure~\ref{fig:epsf}.
Lower S/N lines therefore result in maps with lower effective spatial resolution, but the maps do not become noisy as they would for a per-spaxel approach.
In contrast, the per-spaxel and binned approaches have a fixed spatial scale, and the maps become increasingly noisy as the S/N decreases.
This highlights that our approach is complementary to the typical per-spaxel approach, rather than a drop-in replacement.

The effective spatial resolution or lengthscale appears to behave similarly to a real physical resolution, in that the smoother maps appear as if they were convolved with a larger effective PSF.
For the radial velocities specifically, this effect appears to lead to spatially compact but spectrally sharp features being ``smeared'' in the lower S/N maps.
Comparing the \Halpha and \Hgamma maps, we see that the large $>12 \, \mathrm{km \, s^{-1}}$ features in the central region of the \Halpha map are smoothed into weaker $\sim 8-10 \, \mathrm{km \, s^{-1}}$ features in the \Hgamma map.
This is exactly what would be expected if the models for lower S/N lines have an effectively larger PSF.
This phenomenon is well known instrumentally (e.g. ``beam smearing'' in radio interferometry; \citealt{cotton1989}), but here arises from the structure of the model.
While this means that small-scale features will be lost for lower S/N lines, it also means that the model does not produce spurious small-scale features that are not supported by the data.

In Section~\ref{sec:lvm_model}, we stated that we fixed the covariance kernel parameters during optimisation. 
This may seem counterintuitive, since one might expect them to control the lengthscale of the inferred maps.
However, these parameters control the PSD of the prior, which is subsequently modified by the data to give the \emph{effective} (posterior) PSD and length scale, see Eq.~\eqref{eq:res_scaling}.
This is where the effective resolution comes from: the trade-off between the prior preferring sparse (in frequency space, not L1 sparsity) solutions with the Fourier coefficients closer to zero, and the likelihood preferring non-zero values in order to better explain the data.

As pointed out by \citet{gonzalez-gaitan2019}, this type of approach where information is shared between statistically independent (in terms of noise process) spaxels can result in more accurate predictions than many independent analyses, and is the result of so-called ``shrinkage'' \citep[e.g.][]{gelman2013}, causing the pooling of information across spaxels.
This is a well-known effect in the statistics literature, and remarkably works even in the case where correlations do not exist in the underlying data, i.e. nearby spaxels are not more similar than distant spaxels \citep[Stein's paradox, see][]{stein1956,efron1973}.

Strictly, the effective resolution is actually spatially-dependent, since the S/N varies across the nebula.
The same map can have both high and low effective resolution regions, depending on the local S/N.
This does not violate our stationarity assumption, which applies to the model prior GP covariance, not the posterior mean or MAP.
The lengthscale estimates we presented in Figure~\ref{fig:epsf} are therefore only global averages (see Appendix~\ref{sec:effective_psd}).

\subsection{Practical strengths and current limitations}

Some general strengths of our approach are as follows.
The model can contain any combination of per-spaxel and continuous components, allowing for flexibility and disentangling of different signals.
For example, we included a per-pointing flux calibration and a per-spectrograph wavelength calibration, which allowed us to infer and correct these systematics directly from the data, using only the assumption that the true fluxes and velocities should not have discontinuities across the sky.
Missing data are handled naturally: there is no requirement for interpolation or imputation, since we do not require a regular grid as input.
The model predictions then exist at all points on the sky spanned by the data.
It is also not necessary to subtract stellar flux before fitting the nebular emission; spaxels with bright stars can just be excluded.

That said, the results presented here are intended as a first demonstration of the method applied to real LVM data, and are not science-ready data products.
We did not correct for extinction, treat the stellar contribution in detail, or quantify uncertainties, and for these reasons we did not provide a comparison to the LVM DAP \citep{sanchez2024}. 
Residual artefacts remain, especially in the faintest \ion{C}{2} maps, and dispersion maps are limited by LSF scatter. 

LVM is spatially undersampled, in that the PSF is much smaller than the size of the fibres \citep{herbst2024}.
For oversampled instruments the challenge of disentangling instrument PSF from astrophysical structure would require care, since GP kernels can suffer identifiability issues if not sufficiently different from the PSF, unless the PSF is accurately characterised \citep{tobar2023}.

\subsection{Final steps toward science-ready maps} \label{sec:final_steps}

We did not present line \emph{ratios}, which are the primary diagnostic used for nebular physical conditions and abundances \citep[e.g.][]{kewley2019}.
Naively taking ratios of our flux maps would yield biased results, since lines of different S/N have different effective spatial resolutions. 
In future work, we will address this by jointly modelling multiple lines and including the ratio itself as a component of the model, yielding unbiased ratio maps and potentially higher precision by sharing information across lines as well as spaxels.

We did not attempt to infer the GP kernel parameters, but instead adopted values that produced plausible results across all lines ($l=0.1$ in the model domain, $s=1.0$ in scaled units).
The only adjustment was to increase the variance of the dispersion component slightly, as the default choice ($1.0$) yielded maps that were smoother than expected when compared to the \Halpha analysis in the science overview paper. 
In principle, the kernel parameters should be inferred from the data and allowed to vary between lines, since the prior exerts a stronger influence in the low-S/N regime. 
However, joint optimisation of the objective over both the line intensities and kernel parameters proved unstable, with solutions running to extreme values. 
Cross-validation could provide a pragmatic, data-driven way to tune the kernel parameters \citep[e.g.][]{browne2000}, though at the cost of discarding their uncertainty.

Finally, we did not quantify uncertainties on the model parameters or the line maps.
This is a crucial step toward science-ready data products, but non-trivial for a non-linear hierarchical model with hundreds of thousands of parameters. 
Straightforward re-sampling or bootstrap methods are unattractive because the hierarchical and strong correlation structure in our model would yield very underestimated uncertainties \citep{davison1997,vaart1998}.
Ultimately, we need to characterise, at least approximately, the joint posterior over all parameters, including the GP kernel parameters.
This is challenging due to the very high dimensionality of the model.
While problems the size of the LVM Rosette data for single emission line models may be tractable with gradient-based Markov Chain Monte Carlo samplers \citep{neal2011, hoffman2014}, at least with fewer modes and access to a modern GPU, this will become infeasible for larger datasets and multi-line models.
We therefore likely require a bespoke approach utilising scalable variational inference \citep[e.g.][]{blei2016,zhang2021,knollmuller2019}, normalising flows \citep[e.g.][]{kobyzev2019}, or other approximate methods appropriate for high-dimensional, non-linear posteriors.
As such, we leave this to future work.

\section{Conclusions} \label{sec:conclusion}

We introduced a spectrospatial forward-modelling approach for IFU spectroscopy that models flux, velocity, and dispersion as continuous fields on the sky, using a fast Gaussian process representation to share information between neighbouring spectra. 
The method adapts its effective resolution to the S/N, retaining fine structure for bright lines and pooling over larger scales when the signal is weak.
We presented a first application to 19 LVM pointings of the Rosette Nebula, demonstrating that it produces seamless, physically plausible maps, for both strong and very weak emission lines.
In particular, we recover spatially continuous and resolved maps of metal recombination lines, which are typically too faint to be spatially resolved in \HII regions with other methods.
Per-pointing and per-spectrograph calibration systematics are also able to be inferred directly from the data.
By recovering robust morphology and kinematics where per-spaxel methods would fail or underperform, this framework extends the practical reach of IFU surveys into the very low-S/N regime. 
The implementation is open-source, computationally efficient, easily extensible, and ready to apply to other instruments and science cases.

\appendix

\section{Effective resolution} \label{sec:effective_psd}

\subsection{Linear theory scaling}

We introduced the effective PSD in Section~\ref{sec:adaptive_res} where we used the equivalent kernel picture \citep{sollich2004}, based on linear filter theory, to qualitatively understand the adaptive resolution property of our models.
From Eq.~\eqref{eq:res_scaling}, we see that the denominator $\tilde{k}(\boldsymbol{\omega}) + N_0$ explicitly defines the trade-off between noise and prior in the effective PSD $\tilde{h}(\boldsymbol{\omega})$, and so the resultant ePSF.
We define a critical frequency $\lVert\omega\rVert_{\rm crit}$ where these contributions are roughly equal
\begin{align}
    \tilde k (\lVert\boldsymbol{\omega}\rVert_{\rm crit}) \approx N_0. \label{eq:crit_def}
\end{align}
The 2-dimensional Matern-3/2 kernel in Eq.~\eqref{eq:matern} has PSD \citep[e.g.][]{rasmussen2006}
\begin{align}
    \tilde{k} (\boldsymbol{\omega}) &= s^2 \frac{\pi \, 3^{5/2}}{\ell^3} \left( \frac{3}{\ell^2} + \lVert\boldsymbol{\omega}\rVert^2\right)^{-5/2}, \label{eq:matern_psd}
\end{align}
where $\lVert\boldsymbol{\omega}\rVert$ is the magnitude of the spatial angular frequency vector $\boldsymbol{\omega} = [\omega_\alpha, \omega_\delta]$, or the radial angular frequency.
This is the frequency that our prior ``sees'', given that we use the Euclidean distance metric Eq.~\eqref{eq:metric}.
Thus the kernel PSD scales as
\begin{align}
    \tilde{k} (\boldsymbol{\omega}) \propto \lVert\boldsymbol{\omega}\rVert^{-5}, \label{eq:matern32_psd_scale}
\end{align}
given fixed kernel parameters. With Eq.~\ref{eq:crit_def} we find scaling
\begin{align}
    \lVert\boldsymbol{\omega}\rVert^{-5} \propto N_0,
\end{align}
or equivalently, for a characterstic (spatial) wavelength $\lambda_{\rm crit} \propto \lVert \boldsymbol{\omega} \rVert_{\rm crit}^{-1}$
\begin{align}
    \lambda_{\rm crit} \propto N_0^{1/5}. \label{eq:lambda_N0}
\end{align}

The spectrally integrated S/N for each spaxel $i$, which we denote ${\rm SNR}_i$, is given by
\begin{align}
    {\rm SNR}_i = \frac{\sum_{j}^n F_{ij} \sigma_{ij}^{-2}}{\sqrt{\sum_{j}^n \sigma_{ij}^{-2}}} \label{eq:spectral_int_snr}
\end{align}
where $i$ is an index over spaxels and $j$ is an index over $n$ (spectral) wavelengths.
Assuming the uncertainties are roughly constant over the wavelength range such that $\sigma_i = \sigma_{ij}$, we find
\begin{align}
    {\rm SNR}_i = \frac{\sigma_i^{-1} \sum_j^n F_{ij}}{\sqrt{n}} \propto \sigma_i^{-1}.
\end{align}
Recall that $N_0$ is the variance of the white noise process in the spatial domain.
The value of $N_0$ for spaxel $i$ is then
\begin{align}
    N_{0,i} = \sigma_i^2,
\end{align}
and so
\begin{align}
    N_{0,i} &\propto {\rm SNR}_i^{-2}. \label{eq:sigma_N0}
\end{align}
Combining Eq.~\eqref{eq:lambda_N0} and Eq.~\eqref{eq:sigma_N0}, we obtain
\begin{align}
    \lambda_{\rm{crit},i} \propto {\rm SNR}_i^{-2/5}.
\end{align}
This scaling still holds taking the median ${\rm SNR}$ and $\lambda_{\rm{crit}}$ over spaxels, since the scaling is monotonic and medians are order-preserving
\begin{align}
    \left< \lambda_{\rm crit} \right> \propto \left< {\rm SNR} \right>^{-2/5},
\end{align}
where $\left< \cdot \right>$ denotes the median over spaxels.
Finally, we associate the critical wavelength with the effective PSD lengthscale $\ell_{\rm eff}$.
This follows from the fact that we defined $\lambda_{\rm crit}$ as the spatial wavelength at which the data and prior contribute equally to the effective PSD.
Thus
\begin{align}
    \ell_{\rm eff} \propto {\rm SNR}^{-2/5},
\end{align}
where we have dropped the notation for the median but strictly these quantities are still averages over spaxels.

From this picture we see that in general, $\ell_{\rm eff}$ is actually a local quantity that adapts to the S/N nearby spatially.
That is, in the same model $\ell_{\rm eff}$ could be very short (high resolution) in a high S/N region of the sky, and very long (low resolution) in a low-S/N region of the sky.
Strictly, there is then no global $\ell_{\rm eff}$ that sets the exact resolution for the entire spatial extent of any of the maps, unless the S/N is identical across all spaxels, which is not true even with identical measurement uncertainties since the signal varies.
Instead, the calculated $\ell_{\rm eff}$ values should be treated as an average, but hopefully representative, lengthscales.
Empirically, this seems to be reasonable as the expected scaling roughly holds under this assumption, see Section~\ref{sec:resolution_results} and Figure~\ref{fig:epsf}.

\subsection{$\ell_{\rm eff}$ and ${\rm SNR}$ estimates}

We estimated $\ell_{\rm eff}$ by estimating the effective PSD directly from the Fourier coefficients $X_l$ for each GP component.
We assume that the Fourier coefficients are noisy realisations drawn from
\begin{align}
    X_l \sim \mathcal N \left(0, \tilde{h}(\boldsymbol{\omega}_l) \right), \label{eq:ePSD_sampling}
\end{align}
where the variance for mode $l$ is given by the effective PSD $\tilde h$ at frequency $\boldsymbol{\omega}_l$.
We also assume that the form of $\tilde h$ can be well approximated by the Matern-3/2 PSD in Eq.~\eqref{eq:matern_psd}, meaning that we need to estimate both $s$ and $\ell$.
We estimated $s$ by binning $\lvert X_l \rvert^2$ as a function of $\lVert \boldsymbol{\omega}_l \rVert$, with 30 evenly spaced frequency bins between 0 and 50.
We then took the value of $s$ as the value in the first frequency bin $\lVert \boldsymbol{\omega}_l \rVert \in [0, 5/3]$.
To estimate $\ell$, we then simply found the maximum likelihood estimate $\hat{\ell}$ given the $s$ estimate.
Explicitly, we minimised the negative log-likelihood
\begin{align}
    \hat{\ell} = \argmin_\ell \left[0.5 \sum_l \left( \frac{X_l^2}{\tilde{h}(\boldsymbol{\omega}_l)} + \log{\tilde{h}(\boldsymbol{\omega}_l)} \right)\right],
\end{align}
using the \codesnip{brent} method in \codesnip{scipy.optimize.minimize\_scalar} \citep{2020SciPy-NMeth}.
We then set $\ell_{\rm eff} = \hat{\ell}$.

We calculated the spaxel-median spectrally integrated S/N for each line using the measured fluxes $F_{ij}$ after subtracting the best-fit per-spaxel offsets $\Gamma(i)$ to remove continuum.
These were used in conjunction with the measurement uncertainties $\sigma_{ij}$ in Eq.~\eqref{eq:spectral_int_snr}, before taking the median over spaxels.

\section{Wavelength and flux calibration model} \label{sec:calibration}

The current version of the LVM DRP has a known issue in the wavelength calibration of the blue arm of the spectrographs, where there is a systematic difference across the three spectrographs.
If not accounted for, this dominates the kinematics and leads to a red-blue pattern correlated with footprint of each spectrograph.
This is true for both per-spaxel approaches, but also for our models, where the GP component ends up dominated by trying to account for this difference across the spectrographs for lines in the blue end.
We therefore included a simple correction that assumes that each spectrograph has some unknown shift in wavelength that is the same for all fibres fed to that spectrograph.
We also suspected that this shift would be the same for all pointings, and potentially also for all wavelengths in the blue arm.
To test this, we initially fit for calibration factors as described in Section~\ref{sec:methods}, but we allowed them to be different in each spectrograph and in each pointing.
We tested this by fitting four brighter lines in the blue arm, H$\delta$, \Hgamma, \Hbeta and \OIIIsel, and then plotting the difference in the fitted calibration factors between pairs of spectrographs as a function of pointing.

\begin{figure*}
    \centering
    \includegraphics[width=0.8\textwidth]{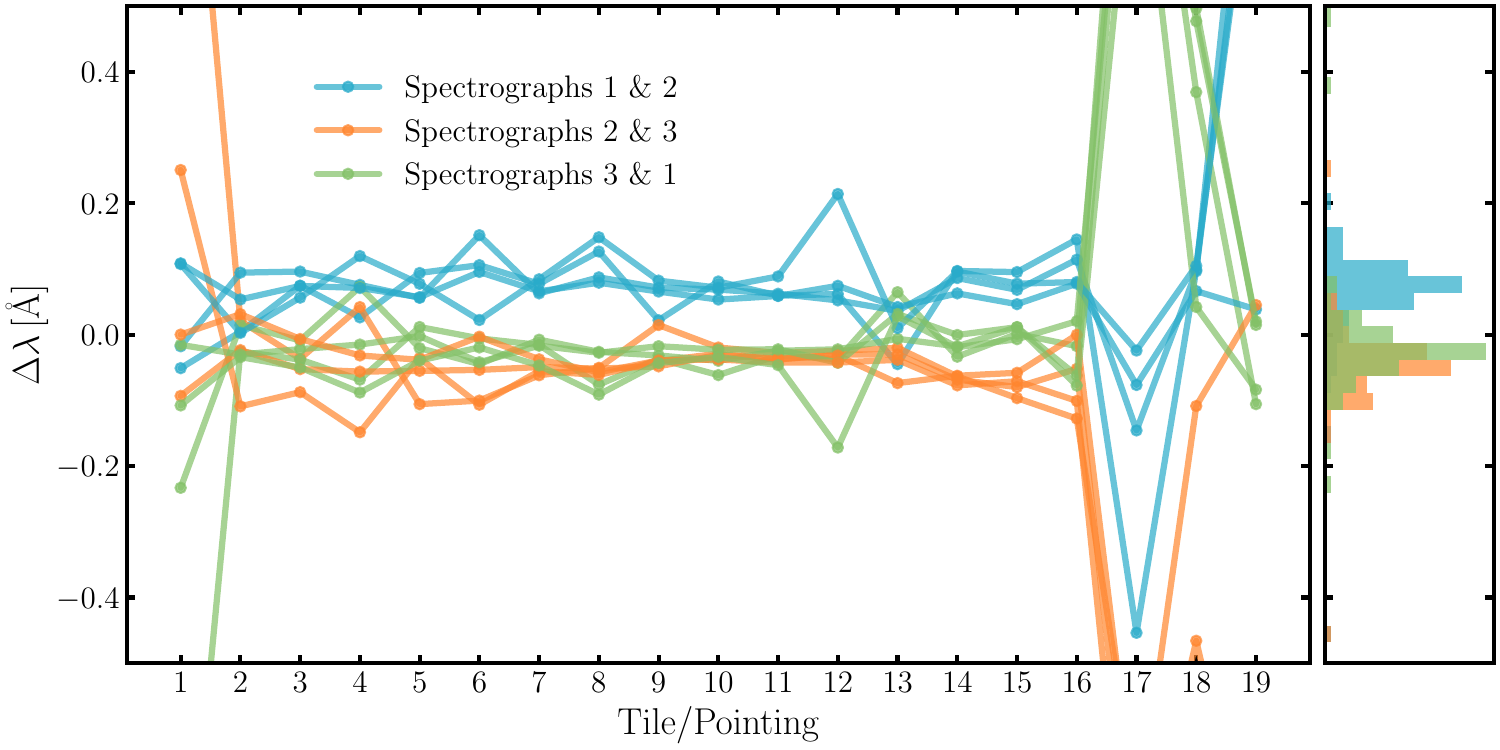}
    \caption{Difference in fitted wavelength calibration factor between pairs of spectrographs in angstroms as a function of pointing. The calibration factors were allowed to vary per pointing and per spectrograph, and fit for four bright lines in the blue arm of the LVM. Each pair of spectrographs is represented by one colour, and the four lines of each colour correspond to the different lines H$\delta$, \Hgamma, \Hbeta and \OIIIsel. These results were intended to test if the wavelength correction required is uniform across the wavelength range and the same for each pointing. We conclude that this is the case as the different coloured lines are both tightly clustered, indicating consistency between lines/wavelengths, and horizontal, indicating uniformity across pointings.}
    \label{fig:v_cal}
\end{figure*}

The results from the test are shown in Figure~\ref{fig:v_cal}.
We see that the calibration factors are tightly clustered for each pair of spectrographs, indicating that indeed the calibration factors are consistent across the different lines and so across the wavelength range in the blue arm.
Additionally, the lines are essentially horizontal, and so it's reasonable to assume that the factors are also the same across pointings.
At pointings 1, 17, 18 and 19 the lines do diverge wildly, but these corresponds to physical locations on the sky where the emission becomes very faint, and so we assumed that these are ignorable.
At these locations the redshift/blueshift of the lines also becomes very large, and so it is possible that the model prefers to use the calibration factors to account for these large shifts rather than the GP component.

For the results for blue-arm lines in main text, we therefore enforced the wavelength calibration factors to be the same for all pointings, and to be parameterised in terms of the difference between the spectrographs such that
\begin{align}
    v_{\mathrm{cal},0} &= \Delta \lambda_0, \\
    v_{\mathrm{cal},1} &= v_{\mathrm{cal},0} - \left( \frac{\Delta \lambda_1}{\mu_{\rm rest}} \right) c \\
    v_{\mathrm{cal},2} &= v_{\mathrm{cal},1} - \left( \frac{\Delta \lambda_2}{\mu_{\rm rest}} \right) c,
\end{align}
where $\Delta \lambda_0$ is an arbitrary shift that is perfectly degenerate with the systemic velocity $v_0$, and so we set it to zero.
The free parameters are therefore $\Delta \lambda_1$ and $\Delta \lambda_2$, which we set to values obtained from a fit to \Hbeta, the brightest line in the blue arm.
If this wavelength calibration factors are not applied for the blue arm, the flux intensity maps are essentially unchanged, the radial velocity maps become dominated by this systematic, and the dispersion maps converge to zero.
As seen in Figure~\ref{fig:vrad_9lines}, the radial velocity maps for the Balmer lines in the blue arm match the morphology of \Halpha very well, and none of the blue-arm lines appear to show the blue-red pattern correlated with the spectrograph footprint, and so we conclude that this approach to wavelength calibration is sufficient.

\begin{figure*}
    \centering
    \includegraphics[width=0.6\textwidth]{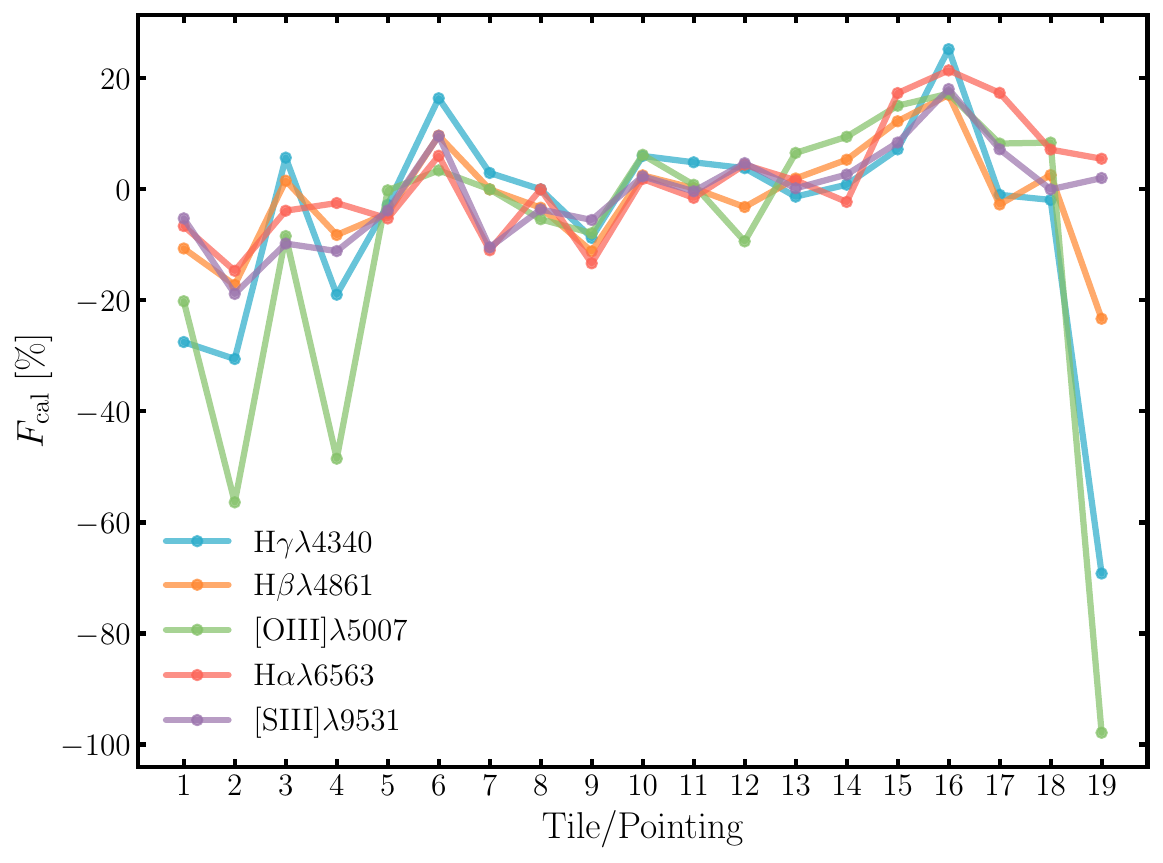}
    \caption{Multiplicative per-pointing flux calibration factors for five bright lines across the whole LVM wavelength range. The different coloured lines represent the calibration factors fit for each emission line. The factors are presented here as a percentage deviation from the median value across pointings obtained for each line. These results were intended to test if the difference in flux calibration between pointings is uniform across the wavelength range. We conclude that this is reasonable as the different coloured lines are for the most part tightly clustered and exhibit the same trends across tiles with some scatter.}
    \label{fig:f_cal}
\end{figure*}

There is also some difference in the flux calibration between pointings that becomes noticable for the brightest lines like \Halpha, appearing as a discrete jump between adjacent pointings in the flux intensity if not accounted for.
We therefore attempted to account for this as part of the model.
We included this as a multiplicative factor that is the same for all fibres in a given pointing, but is allowed to vary between pointings.
To test whether or not the same factors could be applied across the whole wavelength range, we fit for the calibration across five bright lines, with the results shown in Figure~\ref{fig:f_cal}.
We see that at each pointing, the calibration factors derived from lines at different wavelengths are for the most part tightly clustered, indicating that indeed the flux calibration is uniform across the wavelength range.
We therefore used the brightest line \Halpha to inform the flux calibration factors, and then fixed them for all other lines.
We also tested the impact on the results for fainter lines of allowing them to vary for individual lines, and found that the maps did not change significantly.

\section{Line spread function and velocity dispersion} \label{sec:lsf}

In order to produce continuous maps of the velocity dispersion on the sky, it is strictly necessary to differentiate between the instrumental LSF and the intrinsic velocity dispersion of the nebula.
This is because the LSF is only defined at the fibre positions and so it would be meaningless to model it as a continuous function of sky position.
We therefore chose the parameterisation outlined in Section~\ref{sec:methods}, where the total line width in each spaxel is the quadrature sum of the instrumental LSF and the intrinsic velocity dispersion, and the intrinsic dispersion is a strictly positive GP component.

\begin{figure*}
    \centering
    \includegraphics[width=0.9\textwidth]{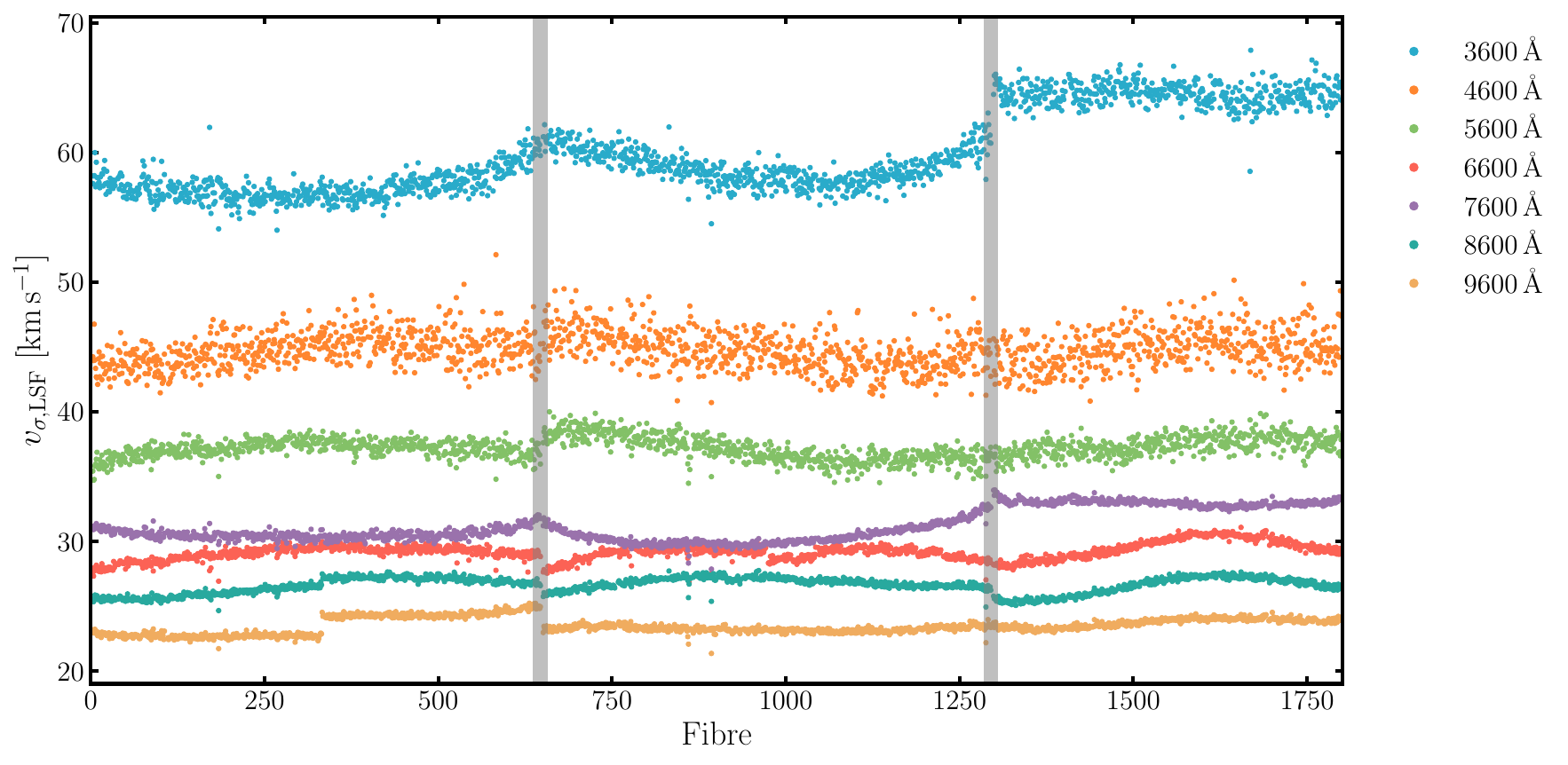}
    \caption{The LSF for a single LVM pointing of Rosette as calculated by the DRP, in units of $\mathrm{km \, s^{-1}}$ and shown as a function of fibre number. These are shown for 7 different wavelengths, represented by the different colours, spanning the whole range covered by LVM. The vertical grey lines represent breaks between the three spectrographs in that the first $\sim$600 fibres are fed to one spectrograph, the next $\sim$600 to the second spectrograph, and the final $\sim$600 to the third spectrograph. These LSF values are calculated by the DRP independently for each fibre. For larger wavelengths, the LSF is much narrower in velocity units, and so contributes less to the total line width in velocity space, allowing the model to have an increased velocity resolution for larger wavelengths. Similarly, the scatter in the LSF values explains the better spatial resolution of the dispersion maps for large wavelengths, as it sets the floor for the smallest resolvable difference in dispersion since the LSF varies randomly between spaxels around the true value, while the dispersion in our models is forced strictly positive and to vary smoothly.}
    \label{fig:lsf}
\end{figure*}

This approach comes with some caveats, namely that the sensitivity, and so probably also the achieved spatial resolution, of the resultant dispersion maps is dependent on the characterisation of the LSF.
Figure~\ref{fig:lsf} shows the LSF for a single pointing of Rosette, as calculated by the LVM DRP, in units of $\mathrm{km , s^{-1}}$ and plotted against fibre number. 
The scatter in the derived LSF is much larger at short wavelengths, which sets a floor in the dispersion maps: because our model enforces the dispersion to be positive and smoothly varying across the sky, deviations smaller than the LSF scatter cannot be resolved.

Some of the scatter is real LSF variation, caused by actual instrumental differences between fibres (for example in positioning).
An additional contribution arises from measurement noise, randomly distributed around the true value.
Outside of this scatter, the LSF appears to vary continuously with fibre number, reflecting the ordering of fibres along the pseudo-slit, with some discontinuities at the breaks between spectrographs.
This suggests that a possible way to refine the LSF measurement for use in spectrospatial approaches like ours would be to fit a continuous function to the LSF, accounting for the breaks between spectrographs.
This assumes that the trends with fibre ID are not just a result of the DRP processing, but rather reflect the true LSF.
It is also unclear how to separate the real scatter from the measurement scatter, which would be necessary to avoid biasing the measured LSF.
Since the primary purpose of this paper is to demonstrate the model for flux intensities, we did not attempt this, but it remains a potential way to improve the dispersion maps for analyses particularly interested in the velocity dispersion.

\vspace{5mm}
\software{
    Astropy \citep{astropycollaboration2013,astropycollaboration2022},
    CMasher \citep{vandervelden2020},
    Dask \citep{daskdevelopmentteam2016},
    Equinox \citep{kidger2021},
    JAX \citep{jax2018github},
    JAX-fiNUFFT \citep{jaxfinufft},
    Matplotlib \citep{Hunter:2007},
    NumPy \citep{harris2020array},
    Optax \citep{deepmind2020jax},
    SciPy \citep{2020SciPy-NMeth},
    xarray \citep{hoyer2017xarray}.
}

\section*{Acknowledgments}


TH is supported by an Australian Government Research Training Program (RTP) Scholarship.
The Flatiron Institute is funded by the Simons Foundation.
KK gratefully acknowledges funding from the Deutsche Forschungsgemeinschaft (DFG, German Research Foundation) in the form of an Emmy Noether Research Group (grant number KR4598/2-1, PI Kreckel) and the European Research Council’s starting grant ERC StG-101077573 ("ISM-METALS").
G.A.B. acknowledges the support from the ANID Basal project FB210003.
SFS thanks the support by UNAM PASPA – DGAPA, the SECIHTI CBF-2025-I-236 and the Spanish Ministry of Science and Innovation (MICINN), project PID2019-107408GB-C43 (ESTALLIDOS), projects.
AMS gratefully acknowledges support by the Fondecyt Regular (project code 1220610), and ANID BASAL project FB210003.
JEMD thanks the support of the SECIHTI CBF-2025-I-2048 project “Resolving the Internal Physics of Galaxies: From Local Scales to Global Structure with the SDSS-V Local Volume Mapper” (PI: Méndez-Delgado).
J.G.F-T gratefully acknowledges the grants support provided by ANID Fondecyt Postdoc No. 3230001 (Sponsoring researcher), and from the Joint Committee ESO-Government of Chile under the agreement 2023 ORP 062/2023.

This work was performed in part on the OzSTAR national facility at Swinburne University of Technology. The OzSTAR program receives funding in part from the Astronomy National Collaborative Research Infrastructure Strategy (NCRIS) allocation provided by the Australian Government, and from the Victorian Higher Education State Investment Fund (VHESIF) provided by the Victorian Government. 

Funding for the Sloan Digital Sky Survey V has been provided by the Alfred P. Sloan Foundation, the Heising-Simons Foundation, the National Science Foundation, and the Participating Institutions. SDSS acknowledges support and resources from the Center for High-Performance Computing at the University of Utah. SDSS telescopes are located at Apache Point Observatory, funded by the Astrophysical Research Consortium and operated by New Mexico State University, and at Las Campanas Observatory, operated by the Carnegie Institution for Science. The SDSS web site is \url{www.sdss.org}.

SDSS is managed by the Astrophysical Research Consortium for the Participating Institutions of the SDSS Collaboration, including the Carnegie Institution for Science, Chilean National Time Allocation Committee (CNTAC) ratified researchers, Caltech, the Gotham Participation Group, Harvard University, Heidelberg University, The Flatiron Institute, The Johns Hopkins University, L'Ecole polytechnique f\'{e}d\'{e}rale de Lausanne (EPFL), Leibniz-Institut f\"{u}r Astrophysik Potsdam (AIP), Max-Planck-Institut f\"{u}r Astronomie (MPIA Heidelberg), Max-Planck-Institut f\"{u}r Extraterrestrische Physik (MPE), Nanjing University, National Astronomical Observatories of China (NAOC), New Mexico State University, The Ohio State University, Pennsylvania State University, Smithsonian Astrophysical Observatory, Space Telescope Science Institute (STScI), the Stellar Astrophysics Participation Group, Universidad Nacional Aut\'{o}noma de M\'{e}xico, University of Arizona, University of Colorado Boulder, University of Illinois at Urbana-Champaign, University of Toronto, University of Utah, University of Virginia, Yale University, and Yunnan University.

\bibliography{lvm.bib}{}
\bibliographystyle{aasjournal}

\end{document}